\documentclass[11pt]{article}
\usepackage{amsbsy}
\usepackage{nicefrac}
\usepackage{datetime}
\usepackage{epsf,psfrag}
\catcode`\@=11
\usepackage{cite}

\usepackage{amsmath}
\usepackage{amsthm}
\usepackage{amsfonts}
\usepackage{amssymb}
\usepackage[cp1251]{inputenc}
\usepackage[english]{babel}

\def \be  {\begin{equation}}
\def \ee  {\end{equation}}
\def \ba  {\begin{eqnarray}}
\def \ea  {\end{eqnarray}}
\def \baa {\begin{eqnarray*}}
\def \eaa {\end{eqnarray*}}
\def \bb  {\begin {thebibliography} }
\def \eb  {\end{thebibliography}}
\def \lab #1 {\label{#1}}

\def \matrix #1 {\left(\begin{array}{cc} #1 \end{array}\right)}

\newcommand{\as}{\ifmmode\alpha_{\rm s}\else{$\alpha_{\rm s}$}\fi}
\newcommand{\asbar}{\ifmmode\bar{\alpha}_{\rm s}\else{$\bar{\alpha}_{\rm s}$}\fi}

\font\cmss=cmss12 
\def\inbar{\,\vrule height1.5ex width.4pt depth0pt}
\def\IC{\relax\hbox{$\inbar\kern-.3em{\rm C}$}}
\def\IZ{\relax{\hbox{\cmss Z\kern-.4em Z}}}
\def\IR{{\hbox{{\rm I}\kern-.2em\hbox{\rm R}}}}

\def\IP{{\hbox{{\rm I}\kern-.2em\hbox{\rm P}}}}
\def\II{\hbox{{1}\kern-.25em\hbox{l}}}

\newbox\lett\newdimen\lheight\newdimen\lwidth
\def\ontop#1#2{\setbox\lett=\hbox{#2}\lheight\ht\lett
\multiply\lheight by 12 \divide\lheight by 10\relax%
\lwidth\wd\lett \multiply\lwidth by 8 \divide\lwidth by 10\relax #2\kern-\lwidth%
\raise\lheight\hbox{{$\scriptstyle #1$}}\kern.1ex}

\def\inbar{\,\vrule height1.5ex width.4pt depth0pt}

\def\id{\hbox{{1}\kern-.25em\hbox{\rm l}}}

\def\be{\begin{equation}}
\def\ee{\end{equation}}
\def\lb#1{\label{#1}}
\def\dd{\partial}
\def\p#1{(\ref{#1})}
\def\mR{\mathrm{R}}
\def\mL{\mathrm{L}}
\newfont{\bbd}{msbm10 scaled\magstep1}
\def\C{\hbox{\bbd C}}

\begin{document}

\begin{titlepage}

\vspace*{1cm}

\begin{center}
{\LARGE \bf{Matrix factorization for solutions \\[0.3 cm] of the Yang-Baxter equation}}

\vspace{1cm}

{ \sc D.~Chicherin\footnote{\sc e-mail:chicherin@lapth.cnrs.fr},
      S.~E.~Derkachov\footnote{\sc e-mail:derkach@pdmi.ras.ru} \\
}

\vspace{0.5cm}

\begin{itemize}
\item[$^1$]
{\it Laboratoire d'Annecy-le-Vieux de Physique Th\'eorique LAPTH, CNRS, UMR 5108,
associ\'ee \`a l'Universit\'e de Savoie, B.P. 110, F-74941 Annecy-le-Vieux, France
}
\item[$^2$]
{\it St. Petersburg Department of Steklov Mathematical Institute
of Russian Academy of Sciences,
Fontanka 27, 191023 St. Petersburg, Russia}
\end{itemize}
\vspace{0.5cm}
\begin{abstract}
We study solutions of the Yang-Baxter equation 
on a tensor product of an arbitrary finite-dimensional
and an arbitrary infinite-dimensional representations of the rank one symmetry algebra.
We consider the cases of the Lie algebra $s\ell_2$, the modular double (trigonometric deformation) and the Sklyanin algebra (elliptic deformation). The solutions are matrices with
operator entries. The matrix elements are differential operators in the case of $s\ell_2$, finite-difference operators
with trigonometric coefficients in the case of the modular double, or finite-difference operators with
coefficients constructed out of Jacobi theta functions in the case of the Sklyanin algebra.
We find a new factorized form of the rational, trigonometric, and elliptic solutions,
which drastically simplifies them. We show that they are products of several simply organized matrices
and obtain for them explicit formulae.
\end{abstract}

\vspace{0.5cm}
\medskip

\hfill{\em To Petr Kulish on the occassion of his 70th birthday}

\medskip

\end{center}
\vspace{4cm}

\end{titlepage}

\tableofcontents


\section{Introduction}
\label{SectIntr}
A quantum integrable system corresponds
to each solution of the Yang-Baxter equation
\be \lb{YB}
\mathbb{R}_{12}(u-v) \,\mathbb{R}_{13}(u) \,\mathbb{R}_{23}(v) =
\mathbb{R}_{23}(v) \,\mathbb{R}_{13}(u)\, \mathbb{R}_{12}(u-v) \,.
\ee
Consequently classification of its solutions is of major interest for
mathematical physics.
The linear operator $\mathbb{R}_{ij}(u)$ from Eq. \p{YB} that acts on a tensor product
of two linear spaces and depends on a spectral parameter $u \in \C$
is traditionally referred to as $\mathrm{R}$-{\it matrix}. We prefer to call it
$\mathrm{R}$-{\it operator}, since we will extensively work with infinite-dimensional linear spaces.
The operators in Eq. \p{YB} act on a tensor product of three linear spaces.
Each $\mathrm{R}$-operator acts non-trivially on a pair of spaces denoted by its lower indices,
and it is extended as an identity operator on the remaining space of the triple.

It is well known that solutions of the Yang-Baxter equation can be rather intricate \cite{KuSk,Jim}.
None the less appealing to the Quantum Inverse Scattering Method~\cite{Fad,KuSk1}
one can put forward a reasonable conjecture that they are composite objects having internal structure
and that they are constructed out of elementary blocks.
A more refined statement is that the $\mR$-operator admits factorization, i.e. it is a product of several simpler operators. 
This observation enabled to construct the general 
solution of the Yang-Baxter equation,Eq. \p{YB},
acting on a tensor product of two infinite-dimensional principal series representations of the group $\mathrm{SL}(N,\C)$ \cite{DM09}.
In the case of rank one algebra this result has been 
carried over to trigonometric and elliptic deformations.
The general $\mR$-operators for the Faddeev's modular double and the elliptic modular double has been constructed in
a factorized from in
\cite{CD14} and \cite{DS1}, respectively.

In this note we will deal with finite-dimensional representations.
We will prove that $\mR$-operators for rank 1 algebras
acting on a tensor product of an {\it arbitrary} finite-dimensional and an arbitrary infinite-dimensional
representations admit factorization as well. These solutions of the Yang-Baxter equation, Eq. \p{YB},
can be thought of as generalizations of the quantum Lax operator,
since the fundamental representation in the auxiliary space $\C^2$
is substituted by a higher-spin representation in $\C^{n+1}$.

Let us consider firstly solutions of Eq. \p{YB} that are invariant with respect to the Lie algebra $s\ell_2$.
In the following sections we will consider as well its trigonometric deformation that is the modular
double (along with $U_q(s\ell_2)$) and its elliptic deformation that is the Sklyanin algebra.

The commutation relations between $s\ell_2$ generators are the following
\be \lb{sl2}
[\,\mathbf{S}^{+}\,,\, \mathbf{S}^{-}\,] = 2 \mathbf{S} \;\; ,
\ \  \ \  [\,\mathbf{S}\,,\,\mathbf{S}^{\pm}\,] = \pm \mathbf{S}^{\pm}\,.
\ee
The symmetry restriction implies commutativity of the $\mathrm{R}$-operator and the co-product of the algebra generators
$$
[\,\mathbb{R}_{12}(u)\,,\, \mathbf{S}^{\pm}_{1} + \mathbf{S}^{\pm}_2 \,] = 0\ \; , \ \ \ \
[\,\mathbb{R}_{12}(u)\,,\, \mathbf{S}_1 + \mathbf{S}_2 \,] = 0\,.
$$
The linear spaces the $\mathrm{R}$-operator acts upon are
representation spaces of $s\ell_2$.
We will be concerned with representations of $s\ell_2$ that are Verma modules.
We realize Verma modules in the space of polynomials $\C[z]$.
The generators of $s\ell_2$ are first order differential operators acting on the space of polynomials
and depending on a parameter $\ell \in \C$, which we call {\it spin} of the representation,
\begin{equation}\label{diff}
\mathbf{S} = z\partial -\ell \ ,\ \mathbf{S}^{-} = \partial \ ,\ \mathbf{S}^{+} =
-z^2\partial + 2 \ell z\ .
\end{equation}
At generic $\ell$ the action of the generators \p{diff} on the space $\C[z]$ is irreducible,
and the representation is infinite-dimensional. At (half)-integer spins
$2 \ell = n \in \mathbb{Z}_{\geq 0}$ the representation is reducible, and
a $(n+1)$-dimensional irreducible representation with the basis $\{1,z,z^2,\cdots,z^n\}$
decouples. We prefer to gather the basis monomials in the generating function $(z-x)^n$ which
depends on an auxiliry parameter $x$. Expanding the generating function with respect to $x$
we recover all basis vectors.

An elegant formula for
$s\ell_2$-invariant solutions of the Yang-Baxter equation, Eq. \p{YB},
acting on a tensor product of two representations of arbitrary
spins $s$ and $\ell$ has been derived in~\cite{KRS81,FTT83},
\be \lb{FTT}
\mathbb{R}_{12}(u|s,\ell) =  \mathrm{P}_{12} \frac{\Gamma(u-J)}{\Gamma(u+J)}\,,
\ee
where $J$ is a ``square root'' of the Casimir operator:
$J(J+1) \equiv (\vec{S}_1 + \vec{S}_2)^2$;
the operator $\mathrm{P}_{12}$ swaps the tensor factors:
$\mathrm{P}_{12} \Phi(z_1,z_2) = \Phi(z_2,z_1)$.
The formula \p{FTT} is valid for both finite-dimensional and infinite-dimensional
representations. The operator $J$ is defined rather formally. Thus the formula \p{FTT}
has to be accompanied with a decomposition of the tensor product
of two representations into irreducibles, which are eigenspaces of $J$.

In \cite{KhT} the universal R-matrix for the Yangian double of $s\ell_2$ has been found in a form of a product of three power series in generators $\mathbf{S},\,\mathbf{S}^{\pm}$, Eq. \p{sl2}.
This universal R-matrix taken in the evaluation representation is an alternative to Eq. \p{FTT}.

Here we choose another opportunity. We will obtain a number of explicit formulae for solutions
of the Yang-Baxter equation, Eq. \p{YB}, working with the functional realization of representations, Eq. \p{diff}.
Indeed, the $\mathbb{R}$-operator acting on the space of polynomials of two complex variables $\C[z_1]\otimes\C[z_2]$ takes the form
\be \lb{Rsl2R}
\mathbb{R}_{12}(u|s,\ell)= \mathrm{P}_{12}
\frac{\Gamma(z_{21}\dd_2-2s)}{\Gamma(z_{21}\dd_2 -u -s - \ell)}
\frac{\Gamma(z_{12}\dd_1 + u -s - \ell)}{\Gamma(z_{12}\dd_1-2s)}
\ee
where $z_{ij} \equiv z_i - z_j$. We imply that representations of the form \p{diff} specified by spins $s$ and $\ell$
are realized in the spaces $\C[z_1]$ and $\C[z_2]$, respectively.
The ratio of two gamma functions of operator argument
can be rewritten as an integral operator by means of the Euler integral of the first kind
$$
\frac{\Gamma(z_{12}\dd_1 + a)}{\Gamma(z_{12}\dd_1 + b)}\,\Phi(z_1,z_2)=
\frac{1}{\Gamma(b-a)}\int_0^1 d\alpha \,\alpha^{a-1}(1-\alpha)^{b-a-1}
\Phi(\alpha z_1+(1-\alpha)z_2,z_2)\,.
$$
Let us note that the $\mathbb{R}$-operator in Eq. \p{Rsl2R} is factorized.
The origin and the meaning of this and other similar factorizations has been clarified in \cite{DKK07}.
The equality of $\mathrm{R}$-operators \p{FTT} and \p{Rsl2R} (up to an inessential normalization factor),
provided the functional realization of $s\ell_2$, Eq. \p{diff}, is adopted,
can be checked by a straightforward calculation \cite{DM09}.

The solution \p{Rsl2R} of the Yang-Baxter equation has been constructed in \cite{Der05} for infinite-dimensional representations of Verma module type.
The spins $s$ and $\ell$ are assumed to be generic.
The case of (half)-integer spins demands an additional refinement. Indeed, the limit
$s \to \frac{n}{2}$ in Eq. \p{Rsl2R} has to be calculated carefully, since the divergences
arise in both factors. In~\cite{CDS} it has been shown that at (half)-integer spin
$2s = n \in \mathbb{Z}_{\geq 0}$ the operator \p{Rsl2R} can be restricted to a finite-dimensional
invariant subspace in the first space of the tensor product.
The restricted operator acts on a tensor product of the $(n+1)$-dimensional space
(where spin $s = \frac{n}{2}$ representation is realized) and
an infinite-dimensional space (where spin $\ell$ representation is realized).
In other words it is a $(n+1)\times (n+1)$ matrix, whose entries are differential operators
acting on the space of polynomials $\C[z]$.
In~\cite{CDS} the restriction of the $\mR$-operator has been calculated and
the generating formula for its matrix matrix elements has been found.
More exactly, the $\mathrm{R}$-operator being applied to the
generating function $(z_1-x)^n$ of the finite-dimensional representation
in the first space and to a polynomial $\Phi(z)$ from the second space\footnote{
In order to simplify notations we use $z$ instead of $z_2$.}
gives the following\footnote{Inessential normalization factors in \p{Rsl2R} and \p{redsl2alg} are different.}
\be\lb{redsl2alg}
\mathbb{R}_{12}(u|{\textstyle\frac{n}{2}},\ell) \,(z_1-x)^n \,\Phi(z) =
\ee
$$
= (z-x)^{-u+\frac{n}{2}+\ell} \, (z_1-z)^{u+\frac{n}{2}+\ell+1} \,
\dd_{z}^n
\, (z_1-z)^{-u + \frac{n}{2} - \ell - 1} \,
(z-x)^{u + \frac{n}{2} - \ell} \, \Phi(z)\,.
$$
Expanding both sides of Eq. \p{redsl2alg} in powers of the auxiliary parameter $x$ we
recover matrix elements of $\mathbb{R}_{12}(u|{\textstyle\frac{n}{2}},\ell)$.
If we choose the second spin in Eq. \p{redsl2alg} to be (half)-integer as well $2\ell = m \in \mathbb{Z}_{\geq 0}$,
then we immediately obtain the restriction of the $\mathrm{R}$-operator to a $(m+1)$-dimensional representation in the second space.
Indeed, applying $\mathbb{R}_{12}(u|{\textstyle\frac{n}{2}},{\textstyle\frac{m}{2}})$ to the generating function $\Phi(z) = (z - y)^m$
of the finite-dimensional representation in the second space,
we find the generating function for matrix elements of $\mathbb{R}_{12}(u|{\textstyle\frac{n}{2}},{\textstyle\frac{m}{2}})$,
which is a $(n+1)(m+1)\times (n+1)(m+1)$ matrix solving the Yang-Baxter equation, Eq.~\p{YB}.

The formula \p{redsl2alg} contains in a compact form all matrix elements of the restricted $\mathrm{R}$-operator.
However the matrix form of the restricted $\mathrm{R}$-operator is still rather obscure.
Using the formula \p{redsl2alg} as a starting point,
we will infer an explicit formula for the restricted $\mathrm{R}$-operator as a
matrix of differential operators. Moreover we will see that this matrix is organized very simply
and that it is much more transparent than \p{redsl2alg}.
In order to get accustomed to \p{redsl2alg}
let us consider several examples.

In the case of restriction to two-dimensional space (spin $s = \frac{1}{2}$)
the formula~(\ref{redsl2alg}) gives rise to the quantum $\mathrm{L}$-operator \cite{Fad}.
In order to see it, let us choose the following basis in $\C^2$: $\mathbf{e}_1 = z_1$, $\mathbf{e}_2 = 1$.
In matrix notations $\mathbf{e}_1 = (1,0)$, $\mathbf{e}_2 = (0,1)$.
Equating coefficients by equal powers of the auxiliary parameter $x$ in both sides of
$$
\mathbb{R}_{12}(u-\textstyle\frac{1}{2}|{\textstyle\frac{1}{2}},\ell) \,(z_1-x) \,\Phi(z) =
(z-x)^{-u+\ell+1} \, (z_1-z)^{u+\ell+1} \,
\dd_{z}
\, (z_1-z)^{-u - \ell} \,
(z-x)^{u - \ell} \, \Phi(z)\,
$$
yields the action of the $\mR$-operator on the basis elements $\mathbf{e}_1, \mathbf{e}_2$
\begin{align*}
\mathbb{R}_{12}(u-\textstyle{\frac{1}{2}}|\textstyle{\frac{1}{2}},\ell) \,\mathbf{e}_1 & = \mathbf{e}_1 \, (z \dd - \ell + u)
+ \mathbf{e}_2 \,(-z^2 \dd + 2 \ell z) \,, \\
\mathbb{R}_{12}(u-\textstyle{\frac{1}{2}}|\textstyle{\frac{1}{2}},\ell) \,\mathbf{e}_2 & = \mathbf{e}_1 \, \dd
+ \mathbf{e}_2 \, (u + \ell - z \dd)\,.
\end{align*}
We tacitly assume that both sides of the previous equalities are applied to an arbitrary polynomial $\Phi(z)$.
Thus the matrix of the operator $\mathbb{R}_{12}(u-\textstyle{\frac{1}{2}}|\textstyle{\frac{1}{2}},\ell)$
in the chosen basis is the following
\be \lb{LaxNonFact}
\mathbb{R}_{12}(u-\textstyle{\frac{1}{2}}|\textstyle{\frac{1}{2}},\ell) =
\begin{pmatrix}
u - \ell + z \dd & \dd \\
-z^2 \dd + 2 \ell z & u + \ell - z\dd
\end{pmatrix} =
\begin{pmatrix}
u + \mathbf{S} & \mathbf{S}^{-} \\
\mathbf{S}^{+} & u - \mathbf{S}
\end{pmatrix}\,.
\ee
It does coincide with the $\mathrm{L}$-operator.
The implemented shift of the spectral parameter simplifies the previous formula.
A straightforward calculation enables to check that the $\mL$-operator
is a product of several upper-triangular and lower-triangular matrices
\be \lb{LaxFact}
\mathbb{R}_{12}(u-\textstyle{\frac{1}{2}}|\textstyle{\frac{1}{2}},\ell) =
\begin{pmatrix}
1 & 0 \\ -z & 1
\end{pmatrix}
\begin{pmatrix}
1 & 0 \\ 0 & u_2
\end{pmatrix}
\begin{pmatrix}
1 & \dd \\ 0 & 1
\end{pmatrix}
\begin{pmatrix}
u_1 & 0 \\ 0 & 1
\end{pmatrix}
\begin{pmatrix}
1 & 0 \\ z & 1
\end{pmatrix}
\,,
\ee
where we introduce linear combinations of the spin and the spectral parameter
$$
u_1 \equiv u - \ell -1 \ \,,\ u_2 \equiv u + \ell\,.
$$
The factorization formula \p{LaxFact} is rather natural, since both the
initial infinite-dimensional $\mR$-operator, Eq. \p{Rsl2R},
and its restriction, Eq. \p{redsl2alg}, have the factorized form.
This leads to a reasonable question: does there exist a factorized matrix form of the
$\mR$-operator at (half)-integer spin $s=\frac{n}{2}$ which would be analogous to \p{LaxFact} ?

In order to answer this question let us consider a more intricate example
of spin $s = 1$, i.e. the restriction to $\mathbb{C}^3$ ($n=2$).
Straightforward application of the formula~(\ref{redsl2alg}) yields the following
matrix of the operator $\mathbb{R}_{12}(u-1|1,\ell)$
written in the basis $\mathbf{e}_1 = z_1^2$, $\mathbf{e}_2 = z_1$,
$\mathbf{e}_3 = 1$
$$
\begin{pmatrix}
u^2 + u (2\mathbf{S}-1) + \mathbf{S}(\mathbf{S} - 1) & - u \mathbf{S}^- + \mathbf{S} \mathbf{S}^- & (\mathbf{S}^-)^2  \\
-2 u \mathbf{S}^+ + 2 \mathbf{S}^+ \mathbf{S} & u^2 - u -2 \mathbf{S}^2 + \ell(\ell+1) & -2 u \mathbf{S}^- - 2 \mathbf{S}^- \mathbf{S} \\
(\mathbf{S}^+)^2 & -u \mathbf{S}^+ - \mathbf{S} \mathbf{S}^+ & u^2 - u (2\mathbf{S}+1) + \mathbf{S}(\mathbf{S} + 1)
\end{pmatrix}\,.
$$
The previous matrix entries are represented through the algebra generators, Eq. \p{diff}.
It is easy to check that the matrix can be decomposed in a product of several
more simply organized triangular matrices
\be \lb{spin1}
\begin{pmatrix}
1 & 0 & 0 \\
- 2z & 1 & 0 \\
z^2 & - z & 1
\end{pmatrix}
\begin{pmatrix}
1 & 0 & 0\\
0 & u_2-1 & 0 \\
0 & 0 & u_2 (u_2-1)
\end{pmatrix}
\begin{pmatrix}
1 & \dd & \dd^2 \\
0 & 1 & 2\dd \\
0 & 0 & 1
\end{pmatrix}
\begin{pmatrix}
u_1(u_1-1) & 0 & 0\\
0 & u_1-1 & 0 \\
0 & 0 & 1
\end{pmatrix}
\begin{pmatrix}
1 & 0 & 0 \\
2z & 1 & 0 \\
z^2 & z & 1
\end{pmatrix}.
\ee
Considering more examples we are able to guess the factorization formula
for the matrix of the operator $\mathbb{R}(u-\textstyle\frac{n}{2}|\textstyle{\frac{n}{2}},\ell)$
written in the basis $\mathbf{e}_1 = z_1^{n}\,, \mathbf{e}_2 = z_1^{n-1}\,,\,\ldots\,,
\mathbf{e}_{n+1} = 1$
\be\lb{formula1}
\mathbb{R}_{12}(u-\textstyle\frac{n}{2}|\textstyle{\frac{n}{2}},\ell) = Z^{-1} \, U^{+}(u_2) \,  D \,  U^{-}(u_1) \,  Z \,.
\ee
For the few first triangular matrices $Z$ and $D$ (at spin $s = \frac{1}{2}\,,1\,,\frac{3}{2}\,,2,\cdots$) we obtain 
\be
Z_{(\frac{1}{2})} = \begin{pmatrix}
1 & 0 \\
z & 1
\end{pmatrix} , Z_{(1)} = \begin{pmatrix}
1 & 0 & 0 \\
2z & 1 & 0 \\
z^2 & z & 1
\end{pmatrix} , Z_{(\frac{3}{2})} =
\begin{pmatrix}
1 & 0 & 0 & 0 \\
3z & 1 & 0 & 0\\
3z^2 & 2z & 1 & 0 \\
z^3 & z^2 & z & 1
\end{pmatrix} ,
Z_{(2)} =
\begin{pmatrix}
1 & 0 & 0 & 0 & 0\\
4z & 1 & 0 & 0 & 0\\
6z^2 & 3z & 1 & 0 & 0 \\
4z^3 & 3z^2 & 2z & 1 & 0\\
z^4 & z^3 & z^2 & z & 1
\end{pmatrix} ,\cdots \lb{Z}
\ee
\be
D_{(\frac{1}{2})} =\begin{pmatrix}
1 & \dd \\
0 & 1
\end{pmatrix} , D_{(1)} =\begin{pmatrix}
1 & \dd & \dd^2 \\
0 & 1 & 2\dd \\
0 & 0 & 1
\end{pmatrix} , D_{(\frac{3}{2})} = \begin{pmatrix}
1 & \dd & \dd^2 & \dd^3\\
0 & 1 & 2\dd & 3\dd^2 \\
0 & 0 & 1 & 3\dd \\
0 & 0 & 0 & 1
\end{pmatrix} ,
D_{(2)} = \begin{pmatrix}
1 & \dd & \dd^2 & \dd^3 & \dd^4 \\
0 & 1 & 2\dd & 3\dd^2 & 4\dd^3 \\
0 & 0 & 1 & 3\dd & 6\dd^2 \\
0 & 0 & 0 & 1 & 4\dd \\
0 & 0 & 0 & 0 & 1
\end{pmatrix} ,\cdots \lb{DD}
\ee
Thus one infers immediately the general pattern.
The diagonal matrices $U^{+}(u)$ for the same values of the spin
are the following
\begin{align}
U^{+}_{(\frac{1}{2})} &= \mathrm{diag}(1,u)\;\;\;,\;\;\;
U^{+}_{(1)} = \mathrm{diag}(1,u-1,u(u-1))\,,
\notag \\
U^{+}_{(\frac{3}{2})} &= \mathrm{diag}(1,u-2,(u-1)(u-2),u(u-1)(u-2))\,,
\notag \\
U^{+}_{(2)} &= \mathrm{diag}(1,u-3,(u-2)(u-3),(u-1)(u-2)(u-3), u(u-1)(u-2)(u-3))\,, \cdots \lb{U+}
\end{align}
The eigenvalues of $U^{-}(u)$ are in a reverse order
\begin{align}
U^{-}_{(\frac{1}{2})} &= \mathrm{diag}(u,1)\;\;\;,\;\;\;
U^{-}_{(1)} = \mathrm{diag}(u(u-1),u-1,1)\,,
\notag \\
U^{-}_{(\frac{3}{2})} &= \mathrm{diag}(u(u-1)(u-2),(u-1)(u-2),u-2,1)\,,
\notag \\
U^{-}_{(2)} &= \mathrm{diag}(u(u-1)(u-2)(u-3),(u-1)(u-2)(u-3),(u-2)(u-3),u-3,1)\,\cdots \lb{U-}
\end{align}
The examined examples lead to a transparent factorization pattern expressed by Eq. \p{formula1}.
The factorization formula \p{formula1} offers considerably more explicit description of finite-dimensional solutions
of the Yang-Baxter equation, Eq. \p{YB}, as compared with all other known approaches. The factorization formula \p{formula1}
is equivalent to the generating formula for matrix elements of the restricted $\mR$-operator, Eq.~(\ref{redsl2alg}),
hence it confirms the efficiency of the result (\ref{redsl2alg}) obtained initially in \cite{CDS}.
Let us mention that it would be difficult to guess the factorization formula~\p{formula1} taking as a starting point
the classical result~\p{FTT}.

The quantum Lax operator (also called $\mL$-operator) is a $2 \times 2$ matrix (the fundamental representation
of a rank 1 algebra is two-dimensional), and its matrix entries are linear in generators of the symmetry algebra.
The $\mL$-operator can be factorized in a product of several more simple matrices in the case of $s\ell_2$
symmetry algebra as well as in the case of its trigonometric and elliptic deformations~\cite{BS,KrZa97,DKK07}.
This observation helps a lot in solving the $\mathrm{RLL}$-relation, which imposes severe constraints on
the infinite-dimensional $\mR$-operator~\cite{DKK07,DM09,DS1,CD14} and eventually enables to find
the general solution of the Yang-Baxter equation, Eq.~\p{YB}.
The purpose of this note is to show that more general solutions of Eq. \p{YB}
than the $\mL$-operator can be factorized as well.

In the next Sect. we will prove the factorization formula~(\ref{formula1}).
Besides the reduction formula \p{redsl2alg} for $s\ell_2$
the analogous restrictions (to a finite-dimensional representation) of the general $\mR$-operator
has been obtained in the paper~\cite{CDS} for the 
Lie group $\mathrm{SL}(2,\C)$ and for the modular double of $U_q(s\ell_2)$~\cite{F99}.
In the accompanying paper~\cite{CDS2} the analogous restriction of the general $\mathrm{R}$-operator \cite{DS1}
has been carried out for elliptic deformations of $s\ell_2$, which are the Sklyanin algebra and the elliptic modular double \cite{AA2008}.
The matrix factorization of $\mathrm{SL}(2,\C)$-symmetric $\mathrm{R}$-operators does not differ essentially
from the formula (\ref{formula1}), since finite-dimensional representations of $\mathrm{SL}(2,\C)$
are tensor products of two $s\ell_2$ representations.
So we will not consider factorization for $\mathrm{SL}(2,\C)$.
In Sect. \ref{SecModDub} and \ref{SecDubFact} we deal with the modular double
and obtain counterparts of the results presented in the current Sect.
The trigonometric factorization is given by the formula \p{trigfactor}.
In Sect. \ref{SecSkl} we consider elliptic deformations and find
the elliptic factorization formula \p{factellip}.

\section{Rational factorization}

In this Sect. we will prove the matrix factorization formula
for the restricted $s\ell_2$-invariant $\mR$-operator, Eq. \p{formula1}.
The reduction formula \p{redsl2alg} obtained in~\cite{CDS}
will be a starting point for us. The proof consists of two steps.
Firstly, we rewrite the matrix formula \p{formula1} in an operator form.
Secondly, we act by this operator on the polynomial $(z_1-x)^n \,\Phi(z)$
and transform the result to the form \p{redsl2alg}.

For the sake of simplicity let us consider firstly an example such that the spin $s = 1$.
We are going to rewrite the factorization formula \p{formula1} at $s = 1$ in an operator form.
Recall that the matrix of the operator $\mathbb{R}_{12}(u-1|1,\ell)$ does factorize, Eq. \p{spin1}.
It is constructed out of matrices $Z_{(1)}$, $D_{(1)}$, $U^\pm_{(1)}$ (see Eqs. \p{Z}, \p{DD}, \p{U+}, \p{U-})
\be \lb{Rspin1}
\mathbb{R}_{12}(u-\textstyle\frac{n}{2}|1,\ell) = Z^{-1}_{(1)} \, U^{+}_{(1)}(u_2)
\,  D_{(1)} \,  U^{-}_{(1)}(u_1) \,  Z_{(1)} \,.
\ee
Each matrix factor in the previous formula has an operator counterpart.
The matrices $Z_{(1)}$ and $D_{(1)}$ have a simple exponential form
$$
Z_{(1)} = \exp \left(z \mathbf{D}_{(1)}\right) \ \ ;\ \
D_{(1)} = \mathbf{C}_{(1)}\,
\exp\left(\dd\, \mathbf{D}_{(1)}\right)\,\mathbf{C}_{(1)}\,,
$$
where we introduce the numerical matrices
\be \lb{DC}
\mathbf{D}_{(1)} \equiv \begin{pmatrix}
0 & 0 & 0 \\
2 & 0 & 0 \\
0 & 1 & 0
\end{pmatrix}\ \ ;\ \
\mathbf{C}_{(1)} \equiv \begin{pmatrix}
0 & 0 & 1 \\
0 & 1 & 0 \\
1 & 0 & 0
\end{pmatrix}\,.
\ee
The matrices $U^+_{(1)}$ and $U^-_{(1)}$ are related to each other by the similarity transformation
$$
U^+_{(1)}(u_2) =
\mathbf{C}_{(1)}\,
U^-_{(1)}(u_2)\,\mathbf{C}_{(1)}\,.
$$
Substituting the previous expressions in Eq. \p{Rspin1} and taking into account that
$\mathbf{C}_{(1)}\,\mathbf{C}_{(1)} = \II$ we rewrite the matrix \p{Rspin1} as follows
$$
\mathbb{R}_{12}(u-1|1,\ell) = \exp\left(-z\, \mathbf{D}_{(1)}\right)\,
\mathbf{C}_{(1)}\,U^{-}_{(1)}(u_2)\,\exp\left(\dd\, \mathbf{D}_{(1)}\right)\,\mathbf{C}_{(1)}\,
U^{-}_{(1)}(u_1)\,
\exp\left(z\, \mathbf{D}_{(1)}\right)\,.
$$
The matrices that constitute the previous expression are matrices of some operators in the basis
$\mathbf{e}_1 = z_1^2$, $\mathbf{e}_2 = z_1$,
$\mathbf{e}_3 = 1$:
\begin{itemize}

\item
the lower-triangular matrix $\mathbf{D}_{(1)}$, Eq. \p{DC}, is a matrix of the differential operator
$\partial_{z_1}$ with respect to the given basis;

\item the matrix $\mathbf{C}_{(1)}$, Eq. \p{DC}, is a matrix of the inversion operator
$\hat{\mathrm{C}}_1 \equiv \hat{\mathrm{C}} \otimes \II:\,z_1^k\to z_1^{n-k}$ at $n=2$ with respect to the given basis;

\item the diagonal matrix $U^{-}_{(1)}(u)$
is a matrix of the operator $\frac{\Gamma(z_1\dd_1+u+1-n)}{\Gamma(u+1-n)}$ at $n=2$ with respect to the given basis\,.
\end{itemize}
Straightforward generalization of
the considered example $s = 1$ (i.e. $n=2$) to arbitrary $n$ yields
an equivalent operator form (with respect to the basis $\mathbf{e}_1 = z_1^{n}\,, \mathbf{e}_2 = z_1^{n-1}\,,\,\ldots\,,
\mathbf{e}_{n+1} = 1$) of the matrix factorization formula, Eq. \p{formula1},
\be \lb{Rmat}
\mathbb{R}_{12}(u-{\textstyle\frac{n}{2}}|{\textstyle\frac{n}{2}},\ell) =
\exp\left(-z\, \dd_1\right)\,
\hat{\mathrm{C}}_1\,
\frac{\Gamma(z_1\dd_1+u_2+1-n)}{\Gamma(u_2+1-n)}\,\exp\left(\dd\, \dd_1\right)\,\hat{\mathrm{C}}_1\,
\frac{\Gamma(z_1\dd_1+u_1+1-n)}{\Gamma(u_1+1-n)}\,
\exp\left(z\, \dd_1\right).
\ee
We proceed to the second step of the proof.
We are going to act by the operator \p{Rmat} on the polynomial $(z_1-x)^{n}\,\Phi(z)$ and to verify
that the result does coincide with the formula for matrix elements of the $\mR$-operator, Eq. \p{redsl2alg}.
Thus we act sequentially on $(z_1-x)^{n}\,\Phi(z)$ by the operator factors from Eq. \p{Rmat}.
At the first step we perform the shift $z_1\to z_1+z$,
\be \lb{step1}
\exp\left(z\, \dd_1\right)\,(z_1-x)^{n}\,\Phi(z) =
(z_1+z-x)^{n}\,\Phi(z)= \sum_{k=0}^{n} \frac{n!}{k!(n-k)!}\,z_1^k\,(z-x)^{n-k}\,\Phi(z)\,.
\ee
At the second step we act by the operator
$\frac{\Gamma(z_1\dd_1+u_2+1-n)}{\Gamma(u_2+1-n)}$ on the previous expression.
The action of this operator on $z_1^k$ is equivalent to the substitution $z_1\dd_1\to k $,
\be \lb{step2}
\sum_{k=0}^{n} \frac{n!}{k!(n-k)!}\,
\frac{\Gamma(k+u_2+1-n)}
{\Gamma(u_2+1-n)}\,z_1^k\,(z-x)^{n-k}\,\Phi(z)\,.
\ee
At the third step we act by the operators $\hat{\mathrm{C}}_1$ and $\exp\left(\dd\, \dd_1\right)$
on $z_1^k$ that is present in Eq. \p{step2},
\be \lb{step3}
\exp\left(\dd\, \dd_1\right)\,\hat{\mathrm{C}}_1\,z_1^k =
\exp\left(\dd\, \dd_1\right)\,z_1^{n-k} = (z_1+\dd)^{n-k} =
\sum_{p=0}^{n-k} \frac{(n-k)!}{p!(n-k-p)!}\,z_1^{n-k-p}\,\dd^{p}\,,
\ee
and at the last step we apply $\exp\left(-z\, \dd_1\right)\,
\hat{\mathrm{C}}_1\,
\frac{\Gamma(z_1\dd_1+u_2+1-n)}{\Gamma(u_2+1-n)}$ to $z_1^{n-k-p}$ that is present in Eq. \p{step3},
\be \lb{step4}
\exp\left(-z\, \dd_1\right)\,
\hat{\mathrm{C}}_1\,
\frac{\Gamma(z_1\dd_1+u_2+1-n)}{\Gamma(u_2+1-n)}\,z_1^{n-k-p} =
(z_1-z)^{k+p}\,\frac{\Gamma(u_2+1-k-p)}{\Gamma(u_2+1-n)}\,.
\ee
Finally, gathering the previous expressions, Eqs. \p{step2}, \p{step3}, \p{step4}, we  obtain
$$
\mathbb{R}_{12}
(u-\textstyle\frac{n}{2}|{\textstyle\frac{n}{2}},\ell)
\,(z_1-x)^{n}\,\Phi(z) =
$$
\be
= \sum_{k=0}^{n} \frac{n!}{k!(n-k)!}\frac{\Gamma(u_1+1-n+k)}{\Gamma(u_1+1-n)}
\sum_{p=0}^{n-k} \frac{(n-k)!}{p!(n-k-p)!}\frac{\Gamma(u_2+1-k-p)}{\Gamma(u_2+1-n)}
\,(z_1-z)^{k+p}\,\partial_z^p\,(z-x)^{n-k}\,\Phi(z)\,. \lb{Right1}
\ee
The previous formula is an operator reformulation of the matrix formula \p{formula1}.
More exactly, according to Eq. \p{Right1}, the matrix \p{formula1} is applied to the
$(n+1)$-dimensional vector $(z_1-x)^{n}$ and operator entries of the matrix \p{formula1} act on a polynomial $\Phi(z)$.
In order to complete the proof we need to show that the right hand side of Eq.~\p{Right1} coincides with
the right hand side of Eq.~(\ref{redsl2alg}) where the spectral parameter is shifted $u \to u - \frac{n}{2}$,
\be\lb{key}
\mathbb{R}_{12}(u-\textstyle\frac{n}{2}|{\textstyle\frac{n}{2}},\ell)\,(z_1-x)^{n}\,\Phi(z)  =
(z-x)^{-u_1-1+n}\,(z_1-z)^{u_2+1}\,\partial_z^n\,
(z_1-z)^{-u_2-1+n}\,(z-x)^{u_1+1}\,\Phi(z)\,.
\ee
The proof of the needed identity will be based on the Cauchy's differentiation formula
\be \lb{Cauchy}
\partial_z^p\, F(z) = \frac{p!}{2\pi \textup{i}}
\oint \frac{d\lambda}{(\lambda-z)^{p+1}} F(\lambda)\,,
\ee
where the closed contour around $\lambda = z$ does not encircle
singularities of an analytic function $F(\lambda)$.

Let us consider the right hand side of Eq. (\ref{Right1}). We reduce fractions canceling $(n-k)!$
and change the summation index in the second sum $p \to n-k-p$
$$
\sum_{k=0}^{n} \frac{n!}{k!}\frac{\Gamma(u_1+1-n+k)}{\Gamma(u_1+1-n)}
\sum_{p=0}^{n-k} \frac{1}{p!(n-k-p)!}\frac{\Gamma(u_2+1-n+p)}{\Gamma(u_2+1-n)}
\,(z_1-z)^{n-p}\,\underline{\partial_z^{n-k-p}\,(z-x)^{n-k}\,\Phi(z)}\,.
$$
Further we exploit the integral representation \p{Cauchy} for the underlined factor,
$$
\sum_{k=0}^{n} \frac{n!}{k!}\frac{\Gamma(u_1+1-n+k)}{\Gamma(u_1+1-n)}
\sum_{p=0}^{n-k} \frac{1}{p!(n-k-p)!}\frac{\Gamma(u_2+1-n+p)}{\Gamma(u_2+1-n)}
\,(z_1-z)^{n-p}\,\frac{(n-k-p)!}{2\pi \textup{i}} \times
$$
$$
\cdot
\oint \frac{d\lambda \;(\lambda-x)^{n-k}}{(\lambda-z)^{n-k-p+1}}
\,\Phi(\lambda) = \frac{n!}{2\pi \textup{i}}\oint \frac{d\lambda}{(\lambda-z)^{n+1}}(z_1-z)^{n}(\lambda-x)^{n}\times
$$
\be \lb{lamint}
\cdot
\sum_{k=0}^{n} \frac{\Gamma(u_1+1-n+k)}{k!\Gamma(u_1+1-n)}\,
\left(\frac{\lambda-z}{\lambda-x}\right)^k\,
\sum_{p=0}^{n-k} \frac{\Gamma(u_2+1-n+p)}{p!\Gamma(u_2+1-n)}
\,
\left(\frac{\lambda-z}{z_1-z}\right)^p\,\Phi(\lambda)\,.
\ee
Both summations in the previous formula can be extended freely to infinity.
Indeed, the unwanted terms contain $(\lambda-z)^{m}$, $m \geq n+1$, and,
consequently, disappear being integrated over $\lambda$.
The emerging power series are of binomial type
$$
\left(1-z\right)^{-\alpha} = \sum_{k=0}^{\infty} \frac{\Gamma(\alpha+k)}{k!\Gamma(\alpha)}\, z^k \,,
$$
so the sums in Eq. \p{lamint} can be evaluated explicitly
$$
\frac{n!}{2\pi \textup{i}}\oint \frac{d\lambda}{(\lambda-z)^{n+1}}\,(z_1-z)^{n}\,
(\lambda-x)^{n}\,
\left(1-\frac{\lambda-z}{\lambda-x}\right)^{n-u_1-1}\,
\left(1-\frac{\lambda-z}{z_1-z}\right)^{n-u_2-1}\,\Phi(\lambda)\,.
$$
Further we rearrange the factors in the previous expression and calculate the contour integral
according to Eq. \p{Cauchy}
that produces immediately the desired result \p{key},
$$
(z-x)^{-u_1-1+n}\,(z_1-z)^{u_2+1}\frac{n!}{2\pi \textup{i}}\oint \frac{\mathrm{d}\lambda}{(\lambda-z)^{n+1}}\,
\left(\lambda-x\right)^{u_1+1}\,
\left(z_1-\lambda\right)^{n-u_2-1}\,\Phi(\lambda) =
$$
$$
=(z-x)^{-u_1-1+n}\,(z_1-z)^{u_2+1}\,\partial_z^n\,
(z_1-z)^{-u_2-1+n}\,(z-x)^{u_1+1}\,\Phi(z)\,.
$$
Thus the identity~(\ref{key}) along with the matrix factorization formula
for the operator $\mathbb{R}_{12}(u-\textstyle\frac{n}{2}|{\textstyle\frac{n}{2}},\ell)$, Eq.~(\ref{formula1}), are proven.

\section{Modular double}
\lb{SecModDub}
In this Sect. we consider solutions of the Yang-Baxter equation, Eq.~\p{YB},
that are invariant with respect to the modular double.
The modular double of the quantum algebra $U_q(s\ell_2)$ has been introduced by Ludwig Faddeev
in~\cite{F99}. This algebra is formed by two sets of generators
$\mathbf{E}\,,\mathbf{K}\,,\mathbf{F}$ and
$\widetilde{\mathbf{E}}\,,\widetilde{\mathbf{F}}\,,\widetilde{\mathbf{K}}$.
The standard commutation relations for the generators $\mathbf{E}\,,\mathbf{K}\,,\mathbf{F}$,
which form the quantum algebra $U_q(s\ell_2)$ with the deformation parameter
$q = e^{\textup{i} \pi \tau}$ (we assume that $\tau \in \C \backslash \mathbb{Q}$,
i.e.  $q$ is not a root of unity)
\begin{equation} \label{qsl2}
\begin{array}{c}
[\,\mathbf{E}\,,\,\mathbf{F}\,] = \frac{\mathbf{K}^2 - \mathbf{K}^{-2}}{q-q^{-1}} \;,\;\;\;
\mathbf{K} \,\mathbf{E} = q \,\mathbf{E} \,\mathbf{K} \;,\;\;\;
\mathbf{K} \,\mathbf{F} = q^{-1}\, \mathbf{F}\, \mathbf{K}\,,
\end{array}
\end{equation}
are supplemented by analogous commutation relations for
$\widetilde{\mathbf{E}},\,\widetilde{\mathbf{F}},\,\widetilde{\mathbf{K}}$
with the deformation parameter $\widetilde{q} = e^{\textup{i} \pi / \tau}$.
In addition, the generators
$\mathbf{E}$ and $\mathbf{F}$ commute with $\widetilde{\mathbf{E}}$
and $\widetilde{\mathbf{F}}$;
the generator $\mathbf{K}$ anticommutes with $\widetilde{\mathbf{E}}$ and
$\widetilde{\mathbf{F}}$; $\widetilde{\mathbf{K}}$ anticommutes with
$\mathbf{E}$ and $\mathbf{F}$; $\mathbf{K}$ and $\widetilde{\mathbf{K}}$ commute.

The representation theory of the modular
double has been elaborated in a number of papers,
see for example~\cite{BT02,F99,FKV,Had,Pawelkiewicz:2013wga}
and references therein.
We will use the following parametrization $\tau = \frac{\omega'}{\omega}$, where
$\omega, \,\omega' \in \C$,
$\mathrm{Im}\, \omega >0$, $\mathrm{Im}\, \omega' >0$,
are constrained by $\omega \omega' = -\frac{1}{4}$.
Then
$$
q = \exp\left(\textup{i} \pi \omega' / \omega \right) \;,\;\;\;
\widetilde{q} = \exp\left(\textup{i} \pi \omega / \omega' \right)\,,
$$
so the change $q \rightleftarrows \widetilde{q}$ is equivalent to $\omega \rightleftarrows \omega'$.
We will also profit from the notation $\omega'' = \omega + \omega'$.

Further we deal with realization of the modular double
generators by finite-difference operators
$\mathbf{K}_s = \pi_s(\mathbf{K})\,,
\mathbf{E}_s = \pi_s(\mathbf{E})\,, \mathbf{F}_s = \pi_s(\mathbf{F})$
acting on the space of entire functions rapidly decaying at
infinity along contours parallel to the real axis. The representation $\pi_s$
is parametrized by a complex number $s$, which we call {\it spin}.
The generators have the following explicit form~\cite{BT02,BT06,CD14}
\begin{equation} \label{Gs}
\mathbf{K}_s = e^{-\frac{\textup{i} \pi}{2\omega} \hat p} \;\;\;,\;\;\;\;
\begin{array}{l}
(q-q^{-1})\mathbf{E}_s =
e^{\frac{\textup{i} \pi x}{\omega}} \left[
e^{-\frac{\textup{i} \pi}{2 \omega}\left(\hat p -s - \omega''\right)} -
e^{\frac{\textup{i} \pi}{2 \omega}\left(\hat p -s - \omega''\right)}
\right] \;,\\[0.3 cm]
(q-q^{-1})\mathbf{F}_s =
e^{-\frac{\textup{i} \pi x}{\omega}} \left[
e^{\frac{\textup{i} \pi}{2 \omega}\left(\hat p + s + \omega''\right)} -
e^{-\frac{\textup{i} \pi}{2 \omega}\left(\hat p + s + \omega''\right)}
\right]\,, 
\end{array}
\end{equation}
where $\hat p$ is a momentum operator in the coordinate representation,
$\hat p = \frac{1}{2 \pi \textup{i}}\, \partial_{x}$.
The formulae
for generators $\widetilde{\mathbf{K}}_s\,,
\widetilde{\mathbf{E}}_s\,,
\widetilde{\mathbf{F}}_s$ are obtained by a change $\omega \rightleftarrows \omega'$ in Eq. \p{Gs}.

\bigskip

The noncompact quantum dilogarithm naturally arises
in the representation theory of the modular double.
In the context of quantum integrable systems
it has been found first in \cite{F95}. The properties of this special function
have been thoroughly examined in ~\cite{FKV,Vol05}. We will need not the quantum dilogarithm itself
but another closely related special function defined by the integral
\be \lb{D}
D_a(z) = \exp\left(-\frac{\textup{i}}{2}\int\limits^{+\infty}_{-\infty}
\frac{d\,t}{t}\,\frac{\sin(a t)\cos(z t)}{\sin(\omega t)\sin(\omega^{\prime} t)}\right)\,,
\ee
where the contour goes above the singularity at $t = 0$.
The R-matrix of the Faddeev-Volkov model is expressed through this function~\cite{VF,Bazhanov:2007mh}.
A number of identities for the $D$-function are contained in~\cite{BT06}.
It naturally arises as an intertwining operator of equivalent representations of the modular double.

Further we indicate some basic properties of the $D$-function that we will need.
The function $D_{a}(z)$ is even,
\be \lb{Dev}
D_a(z) = D_a(-z) \ \ ;\ \  D_a(z) D_{-a}(z) = 1 \ \ ; \ \ D_0 (z) = 1 \,.
\ee
From the definition \p{D} we infer that it is symmetric with respect to the change
$\omega \rightleftarrows \omega'$.
The $D$-function satisfies a pair of finite-difference equations of the first order
\be \lb{FunEq}
\frac{D_a(z-\omega')}{D_a(z + \omega')} =
\frac{\cos \frac{\pi}{2 \omega} ( z-a)}{\cos \frac{\pi}{2 \omega} ( z+a)} \;\; ; \;\;
\frac{D_a(z-\omega)}{D_a(z + \omega)} =
\frac{\cos \frac{\pi}{2 \omega'} ( z-a)}{\cos \frac{\pi}{2 \omega'} ( z+a)}\,.
\ee
Consequently $2\omega$ and $2\omega'$ have the meaning of its quasi-periods.

At generic spin $s$ the representation $\pi_s$ is irreducible and infinite-dimensional.
However it is not of a Verma module type, since the representation space
does not contain a highest-weight vector $\Omega(x)$:
$\mathbf{F}_s \,\Omega(x) = 0$, $\widetilde{\mathbf{F}}_s \,\Omega(x) = 0$,
$\mathbf{K}_s \, \Omega(x) = \lambda \,\Omega(x)$,
$\widetilde{\mathbf{K}}_s \, \Omega(x) = \widetilde{\lambda}\, \Omega(x)$.
The representation $\pi_s$ is a deformed analogue of the principal series
representations of the Lie group $\mathrm{SL}(2,\C)$. The situation drastically changes
at spin $s = s_{n,m} \equiv - \omega''- n \omega - m \omega'$, where integers
$n , m \in \mathbb{Z}_{\geq 0}$ enumerate points
of a quarter-infinite lattice on the complex plane  (or a line, for real $\omega/\omega'$).
In this case the representation $\pi_{s_{n,m}}$ is reducible, and
a $(n+1)(m+1)$-dimensional irreducible representation decouples.
Since we will need finite-dimensional representations, let us consider their
structure in more details. The basis of the $(n+1)(m+1)$-dimensional representation
is formed by monomials
$$ 
\widetilde{X}^{n-2k} X^{m-2l}
\;\;\;\;\;\;\text{at}\;\;\;\; k=0,1,\cdots,n\,,\;\;\; l=0,1,\cdots,m\,,
$$
with respect to the variables
\be \lb{Xvar}
X \equiv X(x) = e^{\frac{\textup{i} \pi}{2\omega}x}
\;\;,\;\; \widetilde{X} \equiv \widetilde{X}(x) = e^{\frac{\textup{i} \pi}{2\omega'}x}\,.
\ee
Therefore any finite-dimensional representation of the modular double is a tensor product
of finite-dimensional representations of $U_q(s\ell_2)$ and $U_{\widetilde{q}}(s\ell_2)$.
For our purposes finite-dimensional representations of spin
$s = s_m \equiv - \omega''- m \omega'$, $m \in \mathbb{Z}_{\geq 0}$, will be sufficient.
Thus we will deal with only a half of the modular double.
By means of the $D$-function, Eq. \p{D}, the basis vectors $X^{m-2l}$, $l=0,1,\cdots,m$,
can be arranged in a sole function. Indeed, $D_{m\omega'}(x-y)$ is a generating
function of the $(m+1)$-dimensional representation. It reduces to a linear combination
of exponents by means of certain contiguous relations similar to \p{FunEq},
\begin{align}
D_{m\omega'}(x-y) =
\prod\limits^{m-1}_{l=0}
\left( Y^{-1} X \, q^{\frac{m-1}{2}-l} + Y X^{-1}
\, q^{-\frac{m-1}{2}+l} \right) \ \ , \ \ Y \equiv Y(y) = e^{\frac{\textup{i} \pi}{2\omega}y} \,,
\lb{modGenFun}
\end{align}
where $y$ is an auxiliary parameter.

\section{Trigonometric factorization}
\lb{SecDubFact}

After a brief introduction to the modular double given in the previous Sect. we will consider
solutions of the Yang-Baxter equation, Eq. \p{YB}, that are invariant with respect to this quantum algebra.
The invariance means that $\mR$-operators commute with the co-product of all six generators.
Let us note that the symmetry restriction imposed by $U_q(s\ell_2)$ is not sufficient
to fix completely (up to an inessential normalization factor) the $\mR$-operator.
Initially the R-operator for the modular double
acting on a tensor product of two infinite-dimensional representations $\pi_{s_{(1)}} \otimes \pi_{s_{(2)}}$
was constructed in \cite{BT06} in a form similar to Eq. \p{FTT} where the role of the Euler beta function is played by
the $D$-function, Eq. \p{D}. The remarks made in Sect. \ref{SectIntr} concerning the formula \p{FTT}
are equally valid for this realization of the $\mR$-operator for the modular double.
In~\cite{CD14} relying on the realization \p{Gs} of the algebra generators
a more explicit formula for the $\mR$-operator has been proposed.
There we demonstrated that the $\mR$-operator can be realized as an integral operator,
and it factorizes in a product of four Faddeev-Volkov type $\mathrm{R}$-matrices~\cite{VF}.
In~\cite{Mangazeev:2014gwa,Mangazeev:2014bqa} an explicit hypergeometric formula for the R-matrix of $U_q(s\ell_2)$
acting on a tensor product of two highest-weight representations has been discovered.
In~\cite{Khoroshkin:2014hla} a universal factorization formula for the trigonometric $\mathrm{R}$-operator
has been obtained by means of the universal $\mathrm{R}$-matrix.

In~\cite{CDS} the integral $\mR$-operator for the modular double acting
on a tensor product of two infinite-dimensional representations $\pi_{s_{n,m}} \otimes \pi_{s}$
has been used as a tool to produce finite-dimensional solutions of the Yang-Baxter equation, Eq. \p{YB}.
There the restriction of the $\mR$-operator to a finite-dimensional representation in the
first space at spin $s_{n,m} \equiv  - \omega''- n \omega - m \omega'\,$, $n,\,m \in \mathbb{Z}_{\geq 0}\,$,
has been found. Let us mention that in~\cite{Pawelkiewicz:2013wga} a similar
restriction has been implemented for Racah-Wigner 6j-symbols.

As we have already explained in the precious Sect., we are going deal
with finite-dimensional representations only of spin $s_{m} \equiv - m \omega' - \omega''$.
A generic finite-dimensional representation of spin $s_{m,n}$ can be taken into account as well
without considerable changes in the following reasoning.
The $\mR$-operator acts on the generating function
of finite-dimensional representations at spin $s_{m}$, Eq. \p{modGenFun},
according to the following formula\footnote{Here we implemented a shift of the spectral parameter $u$ as compared with \cite{CDS}.}
\begin{align} \lb{redmodm} &
\mathbb{R}_{12}(u |s_m,s) \cdot D_{m \omega'}(x_{13})\,\Phi(x_2) =
D_{u_2}(x_{12})\times
\\  & \makebox[1em]{}
\cdot \,D_{-u_1+m \omega'}(x_{23})\cdot
D_{m \omega'}(\hat p_{2})\cdot D_{-u_2+m \omega'}(x_{12})\,
D_{u_1}(x_{23})\,\Phi(x_2)\,, \notag
\end{align}
where $x_3$ is an auxiliary parameter of the generating function.
We recall the shorthand notation $x_{ij} \equiv x_i -x_j$.
Instead of the spectral parameter and spin $s$ we prefer another pair of parameters
$$ 
u_1 = u + \frac{s}{2} \;;\;\;\;
u_2 = u - \frac{s}{2} \,.
$$
In these variables the final result will take a more simple form.
$D_{m \omega'}(\hat p_{2})$ from Eq. \p{redmodm} is a finite-difference operator.
It factorizes in a product of $m$ finite-difference operators of the first order due to the formula \p{modGenFun}.

The generating formula \p{redmodm} uniquely specifies the solution of the Yang-Baxter equation, Eq. \p{YB},
that is a $(m+1)\times (m+1)$ matrix with operator entries. According to Eq. \p{redmodm}, the entries are
finite-difference operators of order $m$. The formula \p{redmodm} enables to obtain a more
explicit realization of the restricted $\mR$-operator.
With respect to the basis
\be \lb{trigbas}
\mathbf{e}_j = X_1^{m+2-2j} \  , \ \ \ \ j=1,\ldots,m+1\,,
\ee
where $X_1 = X_1(x_1)$, Eq. \p{Xvar},
the matrix factorization formula for the restricted $\mR$-operator is valid
\be \lb{trigfactor}
\mathbb{R}_{12}(u |s_m,s) = Z \,M(u_2)\, D \,M(u_1)\, Z^{-1}\,.
\ee
The previous formula is a trigonometric counterpart of the rational factorization, Eq.~\p{formula1}.
The matrices $Z$ and $D$ are diagonal
\be \lb{ZDtrig}
(Z)_{kj} = \delta_{kj} \,X_2^{2k-m-2}\ \ ;\ \
(D)_{kj} = \delta_{kj} \,q^{(m-1)(m+2-2k)}\,e^{(m+2-2k)\omega'\partial_2}\,.
\ee
The coordinate $x_2$ is present only in the matrix $Z$, and the momentum operator
$\hat p_2$ is present only in the matrix $D$. The numerical matrix $M(u)$
is given by the following hypergeometric sum
\begin{align}
\left(M(u)\right)_{kj}  =
\sum_{p} \frac{(q^{2};q^{2})_{j-1}\,(q^{2};q^{2})_{m-j+1}
\,q^{(k-p-1)^2+p(p+2-2j)+(j-1)m - \frac{m^2}{2}}}
{(q^{2};q^{2})_{p}(q^{2};q^{2})_{j-1-p}
(q^{2};q^{2})_{k-p-1}(q^{2};q^{2})_{m+2-j-k+p}}
\,U^{2(2p-j-k+2) + m}\,,
\lb{Mtrig}
\end{align}
where the summation over integer $p$ is from $\mathrm{max}(0,k+j-2-m)$ to $\mathrm{min}(k-1,j-1)$;
the q-Pochhammer symbol $(q^2;q^2)_k \equiv (1-q^2)(1-q^4)\cdots(1-q^{2k})$;
$U \equiv U(u) = e^{\frac{\textup{i}\pi u}{2\omega}}$.

In order to illustrate the formula \p{Mtrig} we indicate the first few matrices $M(u)$, $m = 1, 2 ,3$.
Here we use shorthand notations $M^{(m)} = M^{(m)}(u + m)$,
$$
M^{(1)} =
\begin{pmatrix}
U & U^{-1} \\
U^{-1} & U
\end{pmatrix} \ , \ \
M^{(2)} =
\begin{pmatrix}
U^2 & 1 & U^{-2} \\
q+q^{-1} & q U^2 + q^{-1} U^{-2} & q+q^{-1} \\
U^{-2} & 1 & U^2
\end{pmatrix},
$$
$$
M^{(3)} =
\begin{pmatrix}
U^3 & U & U^{-1} & U^{-3} \\
(q^2 + 1 + q^{-2}) U & q^2 U^3 + (1 + q^{-2}) U^{-1} & q^{-2} U^{-3} + (1 + q^2) U & (q^2 + 1 + q^{-2}) U^{-1} \\
(q^2 + 1 + q^{-2}) U^{-1} & q^{-2} U^{-3} + (1 + q^2) U & q^2 U^3 + (1 + q^{-2}) U^{-1} & (q^2 + 1 + q^{-2}) U \\
U^{-3} & U^{-1} & U & U^3
\end{pmatrix}\,.
$$
In the case $m = 1$ the $\mR$-operator being restricted to the fundamental representation
turns into the quantum Lax operator~\cite{BT06}. The factorization of the $\mathrm{L}$-operator
of the XXZ spin chain has been discovered firstly in~\cite{BaSt} in the context of the chiral Potts models.

\vspace{0.3 cm}

The rest of this Sect. is devoted to the proof of the trigonometric factorization formula, Eq. \p{trigfactor}.
We rewrite the finite-difference operator $D_{m \omega'}(\hat p_{2})$ from Eq. \p{redmodm}
as a sum of shift operators
\be \lb{djsum}
D_{m\omega'}(\hat p_{2}) = \sum^{m+1}_{j=1} d_j\, e^{(m+2-2j)\omega'\partial_2}\,.
\ee
An explicit expression for numerical coefficients $d_j$ will not be relevant for a while (see Eq. \p{dk}).
Then we rearrange the factors in Eq. \p{redmodm}. We collect all functions depending on $u_2$ to the
left of the shift operators and all functions depending on $u_1$ to the right of the shift operators,
\begin{align}
\mathbb{R}_{12}(u | s_m ,s) \cdot D_{m\omega'}(x_{13})
\,\Phi(x_2)
= \sum^{m+1}_{j=1} d_{j}\, D_{u_2}(x_{12})\,
D_{-u_2+m\omega'}(x_{12}+(2j-m-2)\omega') \times \notag \\
\cdot \,e^{(m +2 -2j)\omega'\partial_2}\,
D_{u_1}(x_{23}) \, D_{-u_1+m\omega'}(x_{23}+(2j-m-2)\omega')\,\Phi(x_2)\,. \lb{redmodm2}
\end{align}
So the coordinates are present in Eq. \p{redmodm2} only in the form of the function
$D_{u}(x)D_{-u+m\omega'}(x+(2 j-m-2)\omega')$, where $j=1,\ldots,m+1$.
By means of contiguous relations similar to \p{FunEq} one can check that
this function is given by the following finite product
\begin{align}
&D_{u}(x)D_{-u+m\omega'}(x+(2 j-m-2)\omega') =
\nonumber\\
& = U^{m+2-2j}\,q^{j-\frac{m}{2}-1}\,X^m\,
\prod_{k=0}^{m-j}\left(1+q
\,X^{-1}\,U^{-1}\,q^{2k}\right)\,
\prod_{k=0}^{j-2}\left(1+q^{3-2j}
\,X^{-1}\,U\,q^{2k}\right)\,. \lb{DDprod}
\end{align}
Expanding the right hand side of Eq. \p{DDprod} we obtain the following sum
\begin{align}
D_{u}(x)D_{-u+m\omega'}(x+(2j-m-2)\omega') =
\sum^{m+1}_{k=1} d_{jk}(u)\,X^{m+2-2k}\,, \lb{DDsum}
\end{align}
where $d_{jk}(u)$ are some
numerical coefficients, which will be calculated afterwards (see Eq. \p{djk}).

Now we are ready to calculate the matrix of the operator $\mathbb{R}_{12}(u|s_m,s)$
with respect to the basis~\p{trigbas}. We substitute the expansion \p{DDsum}
into Eq. \p{redmodm2} to the left and to the right of the shift operators.
Then we expand both sides of Eq. \p{redmodm2} in powers of $X_3=X_3(x_3)$, Eq. \p{Xvar}.
The generating function $D_{m\omega'}(x_{13})$ can be expanded with respect to the basis \p{trigbas} and
simultaneously in powers of $X_3$ according to Eq.~\p{modGenFun}.
Equating coefficients by powers of $X_3$ in both sides of Eq.~\p{redmodm2}
yields
\begin{align}
&\mathbb{R}_{12}(u|s_m,s) \cdot d_k\,\mathbf{e}_k
\,\Phi(x_2) = \nonumber\\
&= \sum^{m+1}_{i=1} \mathbf{e}_i\,\left(\sum^{m+1}_{j=1}
d_{ji}(u_2)\,X_2^{2i-m-2}\,
d_{j}\,e^{(m+2-2j)\omega'\partial_2}\,
d_{jk}(u_1)
\,X_2^{m+2-2k}\right)\,\Phi(x_2)\,. \lb{preMat}
\end{align}
Matrix entries of the operator $\mathbb{R}$ are coefficients in the expansion of the vector
$\mathbb{R}\,\mathbf{e}_k$ with respect to the basis \p{trigbas}:
$\mathbb{R}\,\mathbf{e}_k = \sum_{i=1}^{m+1} \mathbf{e}_i\,(\mathbb{R})_{ik}$. Consequently
the formula \p{preMat} produces immediately the matrix entries of $\mathbb{R}_{12}(u | s_m ,s)$,
\begin{align}
\left(\mathbb{R}_{12}(u | s_m ,s)\right)_{ik} =
\left(X_2^{-(m+2-2i)}\,d_{pi}(u_2)\,
d_{p}\right)\,\left(\delta_{pj}e^{(m+2-2j)\omega'\partial_2}\right)\,
\left(\frac{d_{jk}(u_1)}{d_{k}}\,X_2^{m+2-2k}\right)\,. \lb{matrfact}
\end{align}
In the previous formula we tacitly imply summation over repeated indices $p,j$;
the matrix entries are presented in the operator form, and
we omit an arbitrary function $\Phi(x_2)$.
Thus the right hand side of Eq. \p{matrfact} is factorized. It is a product of three matrices:
the diagonal matrix containing shift operators is sandwiched between two matrices
made of numerical coefficients $d_{jk}(u)$, $d_{k}$ and coordinates $X_2^{\pm(m+2-2k)}$.
Stripping off the diagonal matrices that contain $X_2$ from the lateral matrices we obtain
the factorization formula
\be \lb{facttrig2}
\mathbb{R}_{12}(u |s_m,s) = Z \,M_2(u_2)\, \overline{D} \,M_1(u_1)\, Z^{-1}\,.
\ee
The diagonal matrix $Z$ is defined in Eq. \p{ZDtrig}. The diagonal matrix $\overline{D}$
is slightly different from $D$ defined in Eq. \p{ZDtrig},
$$
(\overline{D})_{ik} = \delta_{ik} \,e^{(m+2-2k)\omega'\partial_2}\,.
$$
The numerical matrices $M_1(u)$, $M_2(u)$ are
constructed out of the expansion coefficients $d_{jk}(u)$, $d_{k}$,
\be \lb{M1M2}
\left(M_1(u)\right)_{ik} = \frac{d_{ik}(u)}{d_{k}}\ \ ;\ \  \left(M_2(u)\right)_{ik} = d_{k}\,d_{ki}(u)\,.
\ee
The factorization formula \p{facttrig2} is slightly different from Eq. \p{trigfactor}.
In order to recast the formula \p{facttrig2} into \p{trigfactor}, first of all we need to find
the coefficients $d_{jk}(u)$, $d_{k}$ defined by expansions \p{djsum} and \p{DDsum}.
This goal is easily accomplished by means of the q-binomial theorem
\be \lb{q-binom}
(-x;q^2)_m \equiv \prod_{k=0}^{m-1}\left(1+x\,q^{2k}\right) = \sum_{k=0}^{m}
\frac{(q^2;q^2)_m\, q^{k(k-1)}}{(q^2;q^2)_k(q^2;q^2)_{m-k}}\,x^k \,.
\ee
Indeed, the function $D_{m\omega'}$, which produces coefficients $d_j$, Eq. \p{djsum},
is just the product \p{modGenFun} of the type~\p{q-binom}. Consequently,
\be \lb{dk}
d_k = \frac{(q^2;q^2)_m\, q^{(k-1)(k-m-1)}}{(q^2;q^2)_{k-1}(q^2;q^2)_{m-k+1}}\,.
\ee
The coefficients $d_{jk}(u)$, Eq. \p{DDsum}, can be extracted from the product of
two q-binomial sums. We omit the details of the
calculation that results in
\begin{align}
d_{jk}(u)  =
\sum_{p} \frac{(q^{2};q^{2})_{j-1}\,(q^{2};q^{2})_{m-j+1}
\,q^{(k-p-1)^2+p(p+2-2j) + j-\frac{m}{2}-1}}
{(q^{2};q^{2})_{p}(q^{2};q^{2})_{j-1-p}
(q^{2};q^{2})_{k-p-1}(q^{2};q^{2})_{m-j+2-k+p}}
\,U^{2(2p-j-k+2)+m}\,.
\lb{djk}
\end{align}
The summation limits over integer $p$ in the previous formula are the same as in Eq. \p{Mtrig}.

Let us define
$$
\overline{d}_{jk}(u) \equiv d_j\,d_{jk}(u) \,q^{j(m-1) + 1 -\frac{m(m+1)}{2}}\,.
$$
Substituting the explicit expressions for the coefficients \p{dk}, \p{djk} in the definition
of $\overline{d}_{jk}(u)$, one straightforwardly checks that it is symmetric in indices $j,k$:
$\overline{d}_{jk}(u) = \overline{d}_{kj}(u)$. This observation enables us to simplify Eq.~\p{facttrig2}.
We separate the diagonal matrix $\delta_{ik}\,d_k$, move it from $M_2(u)$ towards $M_1(u)$, Eq. \p{M1M2},
and cancel it. Thus instead of a pair of different matrices $M_1(u)$ and $M_2(u)$
a pair of identical matrices $\left(M(u)\right)_{kj} = d_{jk}(u) \,q^{j(m-1) + 1 -\frac{m(m+1)}{2}}$, Eq. \p{Mtrig},
is present in the factorization formula \p{trigfactor}.
The formula \p{trigfactor} is proven.

It would be interesting to relate the factorization formula \p{trigfactor}
to explicit expressions for $\mathrm{R}$-matrices from~\cite{Mangazeev:2014gwa,Mangazeev:2014bqa}
as well as to the universal factorization formula~\cite{Khoroshkin:2014hla}.

\section{Sklyanin algebra and elliptic factorization}
\lb{SecSkl}
In this Sect. we will factorize solutions of the Yang-Baxter
equation, Eq.~\p{YB}, whose symmetry is encoded by the Sklyanin algebra~\cite{skl1}.
The Sklyain algebra is a two-parametric deformation of $s\ell_2$ or
a one-parametric deformation of $U_q(s\ell_2)$. It serves as a dynamical symmetry algebra of the
8-vertex model~\cite{Baxter1}. The four generators $\mathbf{S}^0,\,\mathbf{S}^1,\,\mathbf{S}^2,\,\mathbf{S}^3$
of the algebra respect commutation relations
\begin{align}
\mathbf{S}^\alpha\,\mathbf{S}^\beta - \mathbf{S}^\beta\,\mathbf{S}^\alpha =
\textup{i}\cdot\left(\mathbf{S}^0\,\mathbf{S}^\gamma +\mathbf{S}^\gamma\,\mathbf{S}^0\right)\,,
\notag \\
\mathbf{S}^0\,\mathbf{S}^\alpha - \mathbf{S}^\alpha\,\mathbf{S}^0 =
\textup{i}\,\mathbf{J}_{\beta \gamma}\cdot \bigl(\mathbf{S}^\beta\,\mathbf{S}^\gamma +\mathbf{S}^\gamma\,\mathbf{S}^\beta \bigr)\,,
\lb{SklAlg}
\end{align}
where the triple $(\alpha,\beta,\gamma)$
is an arbitrary cyclic permutation of $(1,2,3)$.
The structure constants
$\mathbf{J}_{\alpha\beta}=
\frac{\mathbf{J}_{\beta}-\mathbf{J}_{\alpha}}{\mathbf{J}_{\gamma}}$,
$\gamma\neq \alpha,\beta$, are expressed through the Jacobi theta functions
(we assume $\eta \in \mathbb{C}$ and $\theta_a(\eta)\neq 0,\, a=1,\ldots,4$)
\be \lb{strcnst}
\mathbf{J}_{1}=\theta_2(2\eta)\theta_2(0)
\theta_2^{-2}(\eta)\ ;\quad
\mathbf{J}_{2}=\theta_3(2\eta)\theta_3(0)
\theta_3^{-2}(\eta)\ ;\quad
\mathbf{J}_{3}= \theta_4(2\eta)\theta_4(0)
\theta_4^{-2}(\eta)\,.
\ee
We adopt shorthand notations $\theta_{a}(z|\tau) \equiv \theta_a(z)$, $a=1,\cdots,4$,
for theta functions with modular parameter  $\tau \in\mathbb{C}$, Im$(\tau)>0$,
$$
\theta_{1}(z|\tau) \equiv \theta_1(z) = -\sum_{n\in\mathbb{Z}}
\mathrm{e}^{\pi \textup{i} \left(n+\frac{1}{2}\right)^2\tau}\cdot
\mathrm{e}^{2\pi \textup{i}
\left(n+\frac{1}{2}\right)\left(z+\frac{1}{2}\right)}\,.
$$
The rest three theta-functions are obtained by shifts of the argument of $\theta_1$ by quasi-period halves
\begin{eqnarray*}
\theta_{2}(z|\tau)=\theta_1(z+{\textstyle\frac{1}{2}}|\tau)\,, \quad
\theta_{3}(z|\tau)=e^{\frac{\pi \textup{i}\tau}{4}+\pi \textup{i} z}\theta_2(z+{\textstyle \frac{\tau}{2}}|\tau)\,, \quad
\theta_4(z|\tau)= \theta_3(z+{\textstyle\frac{1}{2}}|\tau)\,.
\end{eqnarray*}
Besides theta functions $\theta_a(z)$ with modular parameter $\tau$ we will need as well
theta functions with modular parameter $\frac{\tau}{2}$, which we denote as follows
\be \lb{thetahalf}
\bar\theta_3(z) = \theta_3(z|{\textstyle\frac{\tau}{2}}) \;\;,\;\;
\bar\theta_4(z) = \theta_4(z|{\textstyle\frac{\tau}{2}})\,.
\ee
Two types of theta functions are related to each other by the identity
\be \lb{theta1tothetabar}
2\,\theta_1(x+y)\,\theta_1(x-y) = \bar\theta_4(x)\,\bar\theta_3(y)
-\bar\theta_4(y)\,\bar\theta_3(x)\,.
\ee

In the previous Sect. we have seen that the noncompact quantum dilogarithm (more exactly, the $D$-function)
is omnipresent when one deals with representations of the modular double.
In the case of elliptic deformation the same role is played by the elliptic gamma function~\cite{rui}
\begin{equation}
\Gamma(z)\equiv\Gamma(z|\tau,2\eta)\equiv
\prod_{n,m=0}^{\infty} \frac{1-\mathrm{e}^{-2\pi\textup{i}z}
p^{n+1}q^{m+1}}{1-\mathrm{e}^{2\pi\textup{i}z}
p^ nq^m} \;\;\;,\;\;\; p=e^{2\pi\textup{i}\tau}\;\;,\;\; q=e^{4\pi\textup{i}\eta}
\label{egamma}
\ee
where $|p|,|q|<1$. This function possess a number 
of remarkable properties.
We will need the reflection formula
\be \lb{refl}
\Gamma(z) \,\Gamma(-z + 2\eta +\tau) = 1
\ee
and its quasi-periodicity at the shift by $2\eta$,
\be \lb{shift2eta}
\Gamma(z+2\eta) = \mathrm{R}(\tau)\,e^{\textup{i}\pi z}\,\theta_1(z|\tau)\,\Gamma(z)\;\;\;,\;\;\;
\mathrm{R}(\tau) \equiv \frac{p^{-\frac{1}{8}}}
{\textup{i} (p;p)_\infty}\,.
\ee
In the following we extensively use the short-hand notation
$\Gamma(\pm z \pm x):= \Gamma(z+x) \Gamma(z-x) \Gamma(-z+x) \Gamma(-z-x)$.
Various connections between the Sklyanin algebra and elliptic hypergeometric functions
were considered in~\cite{Rains,ros:elementary,ros:sklyanin,AA2008}.

Let us briefly outline some basic facts about representations of the Sklyanin algebra.
It admits a highly nontrivial explicit
realization of generators as first order finite-difference operators
with elliptic coefficients found by Sklyanin in his pioneering paper~\cite{skl2},
\begin{equation}\label{SklyanMod}
\mathbf{S}^a = e^{\pi\textup{i}z^2/\eta}\frac{\textup{i}^{\delta_{a,2}}
\theta_{a+1}(\eta)}{\theta_1(2 z) } \Bigl[\,\theta_{a+1} \left(2
z-g +\eta\right)e^{\eta\partial_z} - \theta_{a+1}
\left(-2z-g+\eta\right)e^{-\eta\partial_z}\Bigl]e^{-\pi\textup{i}z^2/\eta}.
\end{equation}
The operators depend on a parameter $g  \in \mathbb{C}$ called the {\it spin}.
They act on the space of holomorphic functions of $z$.
In Eq. \p{SklyanMod} we use unconventional similarity transformation by means of
$e^{\pm\pi\textup{i}z^2/\eta}$, whose meaning is explained in~\cite{DS1}.
At generic $g$ the representation \p{SklyanMod} is infinite-dimensional and irreducible.
However, for a discrete set of spin values
$g = g_n \equiv (n+1) \eta + \frac{\tau}{2}$, $n \in \mathbb{Z}_{\geq 0}$,
a $(n+1)$-dimensional representation decouples.
The finite-dimensional representation can be realized in the space
$\Theta^+_{2n}$ of even theta functions of order $2n$.
It is formed by holomorphic functions that are even $f(z) = f(-z)$ and
have simple quasi-periodicity properties under the shifts of $z$ by $1$ and $\tau$:
$$
f(z+1) = f(z) \;\; ,\qquad f(z+\tau) =
\mathrm{e}^{-2 n\pi i\tau -4 n\pi i z } f(z)\,.
$$
The action of the generators \p{SklyanMod} at spin $g =g_n$ is invariant and irreducible
on this space. One can easily check that the monomials constructed out of theta functions \p{thetahalf}
form a basis $\{\varphi_{j}^{(n)}(z)\}_{j = 1}^{n+1}$ in the space $\Theta^+_{2n}$,
\be \label{phi}
\varphi_{j+1}^{(n)}(z) =
\left[\bar\theta_3 \left(z\right)\right]^j \,
\left[\bar\theta_4 \left(z\right)\right]^{n-j} \;\; , \;\; j = 0, 1 , \cdots , n\,.
\ee
The elliptic gamma function, Eq. \p{egamma}, enables to combine the basis elements $\varphi_{j}^{(n)}(z)$ into a sole object.
Indeed, $\Gamma\left(\mp z \mp x + g_n\right)$ is a generating function of
the $(n+1)$-dimensional representation, and it depends on an auxiliary parameter $x$.
Owing to Eqs. \p{theta1tothetabar}, \p{refl}, \p{shift2eta} it reduces to a product of linear combinations of
$\bar\theta_3(z)$ and $\bar\theta_4(z)$,
\be \lb{genellip}
c \cdot\Gamma\left(\mp z \mp x + g_n\right) =
\prod_{r = 0}^{n-1}
\left[ \,\bar\theta_3(z) \,\bar\theta_4\left(x+(n-1-2r)\eta\right)
+ \bar\theta_4(z) \,\bar\theta_3\left(x+(n-1-2r)\eta\right) \,\right],
\ee
where an inessential numerical constant
$c = (-2)^{n} \mathrm{R}^{-2n}(\tau) e^{-\frac{i\pi \tau}{2}n}$.
The previous product is equivalent to a linear combination of basis vectors $\varphi^{(n)}(z)$, Eq. \p{phi},
with some coefficients $\psi^{(n)}(x)$ depending on the auxiliary parameter $x$.
The generating function, Eq. \p{genellip}, is invariant under the change $z \rightleftarrows x$,
hence it contains the second natural basis $\{\psi^{(n)}_j(z)\}_{j = 1}^{n+1}$,
$$
c \cdot\Gamma\left(\mp z \mp x + g_n\right) = \sum_{j = 1}^{n+1} \psi_{n+2-j}^{(n)}(x)\, \varphi_{j}^{(n)}(z)
= \sum_{j = 1}^{n+1} \varphi_{n+2-j}^{(n)}(x)\,\psi_{j}^{(n)}(z)\,,
$$
that is formed by products of $\bar\theta_3(z)$ and $\bar\theta_4(z)$ with shifted arguments,
\be \lb{psi}
\psi_{j+1}^{(n)}(z) =  \mathrm{Sym}
\prod_{r=0}^{n-1} \bar\theta_{a_r}\left(z+(n-1-2r)\eta\right) \;\; , \;\; a_r \in \{3,4\} \;\; ,\;\; j = 0, 1 , \cdots , n
\ee
where $\bar\theta_3$ appears $j$ times and $\bar\theta_4$ appears $n-j$ times;
symmetrization $\mathrm{Sym}$ is over indices $\{a_r\}$.

Let us denote a pair of basis \p{phi}, \p{psi} of the $(n+1)$-dimensional space $\Theta^+_{2n}$
by $\{\mathbf{e}_{j}\}_{j=1}^{n+1}$ and $\{\mathbf{f}_j\}_{j=1}^{n+1}$,
$$
\mathbf{e}_{j} = \varphi^{(n)}_j(z)\;\; , \;\;
\mathbf{f}_{j} = \psi^{(n)}_j(z) \;\; , \;\; j = 1 , 2, \cdots , n+1 \,.
$$
At $n = 1$ the representation is $2$-dimensional, the bases coincide
\be \lb{efBasLax}
\mathbf{e}_1 = \mathbf{f}_1 = \bar\theta_4(z) \;\;,\;\; \mathbf{e}_2 = \mathbf{f}_2 = \bar\theta_3(z) \,.
\ee
At higher spins the bases are different. At $n = 2$ the representation is $3$-dimensional, a pair of bases is
$$
\mathbf{e}_1 = \bar\theta^2_4(z)\;,\; \mathbf{e}_2 = \bar\theta_4(z)\bar\theta_3(z)\;,\; \mathbf{e}_3 = \bar\theta^2_3(z) \;;
$$
$$
\mathbf{f}_1 =
\bar\theta_4(z-\eta)\bar\theta_4(z+\eta) \;,\; \mathbf{f}_2 = \bar\theta_4(z-\eta)\bar\theta_3(z+\eta) +
\bar\theta_3(z-\eta)\bar\theta_4(z+\eta) \;,\; \mathbf{f}_3 = \bar\theta_3(z-\eta)\bar\theta_3(z+\eta) \,.
$$

These basic facts about the Sklyanin algebra and its representations will be sufficient for our purposes.
Let us take a look at the corresponding solutions of the Yang-Baxter equation, Eq. \p{YB}.
The symmetry restrictions imposed by the Sklyanin algebra do not allows to fix uniquely
the solution $\mathbb{R}_{12}(u)$ acting on a tensor product of two infinite-dimensional representations
specified by spins $g_{(1)}$ and $g_{(2)}$~\cite{DS1}.
However, more severe restrictions produced by the elliptic double
enable to fix the $\mathrm{R}$-operator unambiguously (up to an inessential constant).
This $\mR$-operator has been constructed in~\cite{DS1} in a form of an integral operator acting
on a tensor product of two arbitrary infinite-dimensional representations of the elliptic double.
The integral kernel of this operator is given by the product of elliptic gamma functions, Eq.~\p{egamma}.
The proof that this integral operator with an elliptic hypergeometric kernel solves the Yang-Baxter
equation is based on a number of sophisticated identities:
the elliptic beta integral evolution formula~\cite{spi:umn,spi:essays},
an integral Bailey lemma~\cite{spi:bailey},
and elliptic Fourier transformation~\cite{spi-war:inversions}.
The elliptic beta integral evolution formula is equivalent to the star-triangle relation~\cite{BS}.
The elliptic double consists of two Sklyanin algebras,
whose structure constants, Eq. \p{strcnst}, are parametrized by $2\eta,\tau$ and $\tau,2\eta$,
so their generators commute or anticommute with each other.
Finite-dimensional representations of the modular double are equivalent (up to a sign)
to a tensor product of finite-dimensional representations of the Sklyanin algebras.
Since we are aimed at finite-dimensional representations and matrix realizations
of $\mR$-operators, we will consider only one of two Sklyanin algebras constituting the elliptic double.

In~\cite{CDS2} the integral $\mR$-operator for the elliptic double has been taken as a starting point
and restrictions of this operator to finite-dimensional representations have been implemented.
In particular, the restriction to a $(n+1)$-dimensional representation in the first space at
spin $g_n \equiv (n+1)\eta +\frac{\tau}{2},\, n \in \mathbb{Z}_{\geq 0}$, has been considered.
The action of the $\mR$-operator on the generating function of the finite-dimensional representation, Eq. \p{genellip},
is given by the formula\footnote{In order to simplify the notation we denote the complex variable $z_2$ by $z$.}
\begin{multline} \lb{redsl2ellip}
\mathbb{R}_{12}(u|g_{n} \,,\,g)\,\Gamma(\mp z_1 \mp z_3 + g_{n})\, \Phi(z) =  \\
=
\frac{\Gamma(\mp z \mp z_3 -\frac{u}{2}+\frac{g_{n}+g}{2})}
{\textstyle\Gamma(\mp z_1\mp z -\frac{u}{2}-\frac{g_{n}+g}{2} + \eta+\frac{\tau}{2})}\,
\mathrm{M}( n \,\eta)\,
\frac{\Gamma(\mp z_1 \mp z -\frac{u}{2}+\frac{g_{n}-g}{2})}
{\Gamma(\mp z \mp z_3 -\frac{u}{2}+\frac{g-g_{n}}{2} + \eta +
\textstyle\frac{\tau}{2})}\,\Phi(z)\,.
\end{multline}
An arbitrary holomorphic functions $\Phi(z)$ belongs to
the second space where a generic spin $g$ representation is realized;
$z_3$ is an auxiliary parameter of the generating function.
The finite-difference operator from the previous formula
\be \lb{Mintw}
\mathrm{M}(n \eta) = \sum_{l = 0}^{n} \beta^{(n)}_l(z) \,e^{(n-2l)\eta\dd_z}
\ee
is an intertwining operator of equivalent representations of the Sklyanin algebra.
For the first time it has been constructed by A. Zabrodin in~\cite{Z}.
In~\cite{DS2} the factorized representation for the intertwiner has been found
$$
\mathrm{M}(n \eta) =
\mathrm{A}_a(n\eta-\eta)\cdots \mathrm{A}_a(\eta) \mathrm{A}_a(0)\cdot
\bar\theta_a^{-n} \left(z\right) \;\;,\;\;\;\;\; a = 3, 4\,.
$$
It is a product of finite-difference operators of the first order
$$
\mathrm{A}_a(g) = e^{\pi \textup{i}\frac{z^2}{ \eta}}\,\frac{1}{\theta_1(2z | \tau)}
\left[ \bar\theta_a \left(z+g+\eta \right)\, e^{\eta \partial_z} -
\bar\theta_a \left(z-g-\eta \right)\, e^{-\eta \partial_z}
\right]\, e^{-\pi \textup{i}\frac{z^2}{ \eta}} \,.
$$
The coefficients $\beta^{(n)}_l(z)$, Eq. \p{Mintw}, can be found in \cite{DS2,Z}.

Expanding \p{redsl2ellip} by means of Eq. \p{genellip} and equating coefficients
on both sides of the formula that accompany the linear independent functions
$\{\phi^{(n)}_j(z_3)\}$ of the auxiliary parameter $x_3$, we obtain the matrix form of
the restricted $\mR$-operator
$$
\mathbb{R}_{12}(u|g_{n} \,,\,g)\,\psi^{(n)}_{j}(z_1) =
\varphi^{(n)}_{l}(z_1) \bigl(\mathbb{R}_{12}(u|g_{n} \,,\,g)\bigr)_{lj}\,
$$
with respect to the pair of bases \p{phi}, \p{psi}:
$\{\mathbf{e}_{j}\}_{j=1}^{n+1}$ and $\{\mathbf{f}_j\}_{j=1}^{n+1}$,
$$
\mathbf{e}_{j} = \varphi^{(n)}_j(z_1)\;\; , \;\;
\mathbf{f}_{j} = \psi^{(n)}_j(z_1) \;\; , \;\; j = 1 , 2 , \cdots , n +1\,.
$$
The matrix elements are finite-difference operators
of the $n$-th order whose coefficients are constructed out of theta functions.
It happens that the matrix form of the restricted $\mR$-operator is more illustrative than Eq. \p{redsl2ellip}.
Indeed, this matrix solution of the Yang-Baxter equation can be factorized as follows
\be \lb{factellip}
\mathbb{R}_{12}(u|g_{n} \,,\,g) = V(u_1,z) \,D(z,\dd)\, \mathbf{C} \, V^{T}(u_2,z) \, \mathbf{C}\,.
\ee
The matrix $D(z,\dd)$ is diagonal, and it is formed by the terms of
the intertwining operator $\mathrm{M}(n \eta)$, Eq.~\p{Mintw},
$$
\left(D(z,\dd)\right)_{lj} = \delta_{lj}\, \beta^{(n)}_{l-1}(z)\,e^{(n+2-2l)\eta \dd_z}\,.
$$
In the numerical matrix $\mathbf{C}$ only the antidiagonal is nonzero:
$\left( \mathbf{C} \right)_{lj} = \delta_{n+2-l , j}$\,. We see that
it is convenient to arrange the spectral parameter $u$
and the spin $g$ in the linear combinations
$$
u_1 = \frac{u + g}{2} \;\;,\;\; u_2 = \frac{u-g}{2}\,.
$$
The matrix $V$ consists of theta functions $\left(V(u,z)\right)_{jl} = V^{(n)}_{jl}(u,z)$
that are specified by the following defining relation
$$
\sum_{j = 1}^{n+1} \varphi_j^{(n)}(x)\,V_{jl}^{(n)}(z,u)  \equiv
\prod_{r = 0}^{n - l}\theta_1\left({\textstyle \pm x + z - u + \frac{g_n}{2} + 2\eta(\frac{n}{2} - l - r)  }\right)
\prod_{r = 2}^{l}\theta_1\left({\textstyle \pm x + z + u - \frac{g_n}{2} + 2\eta( \frac{n}{2}-l + r)  }\right)\,.
$$
In view of Eq. \p{theta1tothetabar}, the function
$V_{jl}^{(n)}$ is a linear combination of theta functions $\bar\theta_3$ and $\bar\theta_4$,
whose arguments are shifted in a certain way. Each monomial contains $j$ times $\bar\theta_4$ and
$n-j$ times $\bar\theta_3$, i.e.
\begin{align}
V_{jl}^{(n)}(z,u) = (-1)^{n+1-j} \,\mathrm{Sym}
\sideset{}{_{r=2}^{l}}\prod\bar{\theta}_{a_{r-1}}
\left({\textstyle \pm x + z + u - \frac{g_n}{2} + 2\eta( \frac{n}{2}-l + r)  }\right)
\times \notag\\
\cdot \,
\sideset{}{_{r=0}^{n-l}}\prod\bar{\theta}_{a_{r+l}}
\left({\textstyle \pm x + z - u + \frac{g_n}{2} + 2\eta(\frac{n}{2} - l - r) }\right),
\notag
\end{align}
where $a_r \in \{3,4\}$.
Let us note that an immediate corollary of the defining relation is
$V_{jl}^{(n)}(-z,u) = V_{j,n+2-l}^{(n)}(z,u)$, i.e.
$V(-z,u) = V(z,u)\,\mathbf{C}$.
The proof of Eq. \p{factellip} follows the line  of reasoning used to prove the factorization formula \p{trigfactor}
for the modular double in Sect. \ref{SecDubFact}. It relies on the properties \p{refl}, \p{shift2eta}
of the elliptic gamma function.

In order to elucidate the formula \p{factellip} we indicate the matrix factors that are
involved in Eq. \p{factellip} at $n=1$ and $n=2$. Diagonal matrices $D_{(n)}$:
\begin{align}
D_{(1)} &= e^{\pi\textup{i}z^2/\eta} \,\frac{1}{\theta_1(2z)} \mathrm{diag}(
e^{\eta \dd} ,  - e^{-\eta \dd} ) \,e^{-\pi\textup{i}z^2/\eta} \;, \notag\\
D_{(2)} &= e^{\pi\textup{i}z^2/\eta}\,
\frac{1}{\theta_1(2z-2)\theta_1(2z)\theta_1(2z+2)}
\mathrm{diag}\left(
\theta_1(2z-2) e^{2\eta \dd} , - \frac{\theta_1(4\eta)}{\theta_1(2\eta)}\theta_1(2z) ,
\theta_1(2z+2) e^{-2\eta \dd} \right) \, e^{-\pi\textup{i}z^2/\eta} \,.\notag
\end{align}
Matrices $V_{(n)}(u)$:
$$
V_{(1)}(u+{\textstyle\frac{\tau}{4}}) = \left(
\begin{array}{cc}
-\bar\theta_3\left(z - u\right) &
-\bar\theta_3\left(z+u \right) \\
\bar\theta_4\left(z - u\right) &
\bar\theta_4\left(z+u\right)
\end{array} \right),
$$
$$
V_{(2)}(u{\textstyle- \frac{\eta}{2}+ \frac{\tau}{4}}) = \left(
\begin{array}{ccc}
\bar\theta_3\left(z - u\right) \bar\theta_3\left(z - u + 2\eta\right) &
\bar\theta_3\left(z - u \right) \bar\theta_3\left(z+u \right) &
\bar\theta_3\left(z + u \right) \bar\theta_3\left(z+u- 2\eta\right) \\
\bar\theta_{\{3}\left(z - u\right) \bar\theta_{4\}}\left(z - u + 2\eta\right) &
\bar\theta_{\{3}\left(z - u \right) \bar\theta_{4\}}\left(z+u \right) &
\bar\theta_{\{3}\left(z + u \right) \bar\theta_{4\}}\left(z+u- 2\eta\right) \\
\bar\theta_4\left(z - u\right) \bar\theta_4\left(z - u + 2\eta\right) &
\bar\theta_4\left(z - u \right) \bar\theta_4\left(z+u \right) &
\bar\theta_4\left(z + u \right) \bar\theta_4\left(z+u- 2\eta\right)
\end{array} \right)\,.
$$
The curly brackets in the second line of the previous formula denote symmetrization with
respect to the theta function indices. Let us recall that at $n = 1$ the bases
$\mathbf{e}$ and $\mathbf{f}$, Eq. \p{efBasLax}, are identical and the restricted $\mR$-operator
coincides with the quantum elliptic Lax operator. The factorization of the elliptic
$\mL$-operator appeared before in \cite{KrZa97}.

\section*{Acknowledgment}

This work is supported by the Russian Science Foundation
(project no. 14-11-00598).




\begin{thebibliography}{99}

\bibitem{Baxter1}
R.~J.~Baxter,~{\it Partition function of the eight-vertex lattice model},
  Ann. Phys.  {\bf 70} (1972), 193--228.	

\bibitem{Bazhanov:2007mh}
  V.~V.~Bazhanov, V.~V.~Mangazeev and S.~M.~Sergeev,
  {\it Faddeev-Volkov solution of the Yang-Baxter equation and discrete conformal symmetry},
  Nucl.\ Phys.\ B {\bf 784} (2007) 234
  [hep-th/0703041].
	
\bibitem{BS}
V.~V.~Bazhanov and S.~M.~Sergeev,
\textit{A master solution of the quantum
Yang-Baxter equation and classical discrete integrable equations},
ATMP {\bf 16} (2012),  65--95, arXiv:1006.0651 [math-ph]

\bibitem{BaSt}	
V.~V.~Bazhanov, Yu.~G.~Stroganov,~{\it Chiral Potts model
as a descendant of the six-vertex model},
J. Stat. Phys. {\bf 59} (1990), 799--817
	
\bibitem{BT02}
A.~G.~Bytsko, J.~Teschner,
{\em R operator, coproduct and Haar measure for the modular double of $U_q(sl(2,R))$},
Commun. Math. Phys.  {\bf 240} (2003), 171--196; math.QA/0208191.

\bibitem{BT06}
  A.~G.~Bytsko, J.~Teschner,
  {\it Quantization of models with non-compact quantum group symmetry:
Modular XXZ magnet and lattice sinh-Gordon model,}
  J. Phys. A {\bf 39} (2006), 12927; hep-th/0602093.

\bibitem{CD14}
  D.~Chicherin, S.~Derkachov,
  {\em The R-operator for a modular double},
  J. Phys. A {\bf 47} (2014), 115203; arXiv:1309.0803 [math-ph].
	
\bibitem{CDS}
  D.~Chicherin, S.~E.~Derkachov and V.~P.~Spiridonov,
  {\it From principal series to finite-dimensional solutions of the Yang-Baxter equation},
  arXiv:1411.7595 [math-ph].
	
\bibitem{CDS2}
  D.~Chicherin, S.~E.~Derkachov and V.~P.~Spiridonov,
  {\it New elliptic solutions of the Yang-Baxter equation},
  arXiv:1412.3383 [math-ph].

\bibitem{Der05}
S.~E.~Derkachev, {\it Factorization of the R-matrix. I},
Zapiski POMI {\bf 335} (2006), 134--163
(J. Math. Sciences {\bf 143} (1) (2007), 2773--2790);
arXiv:math/0503396 [math.QA]

\bibitem{DKK07}
  S.~Derkachov, D.~Karakhanyan, R.~Kirschner,
  {\it Yang-Baxter R operators and parameter permutations},
  Nucl.\ Phys.\ B {\bf 785} (2007) 263
  [hep-th/0703076 [HEP-TH]].

\bibitem{DM09}
S.~E.~Derkachov and A.~N.~Manashov,
{\em General solution of the Yang-Baxter equation with symmetry
group $\mathrm{SL}(n,\mathbb{C})$},
Algebra i Analiz {\bf 21} (4) (2009), 1--94
(St. Petersburg Math. J. {\bf 21} (2010), 513--577).	

\bibitem{DS1}
S.~E.~Derkachov and V.~P.~Spiridonov,
{\em Yang-Baxter equation, parameter permutations, and the elliptic beta integral},
Uspekhi Mat. Nauk {\bf 68} (6) (2013), 59--106
(Russian Math. Surveys  {\bf 68} (6) (2013), 1027--1072);
 arXiv:1205.3520 [math-ph].

\bibitem{DS2} S.~E.~Derkachov and V.~P.~Spiridonov, {\em Finite dimensional
representations of the elliptic modular double}, Theor. Math. Phys., to appear;
arXiv:1310.7570 [math.QA].

\bibitem{Fad} 	
L.~D.~Faddeev,~{\it How Algebraic Bethe Anstz works
for integrable model}, In: Quantum Symmetries/Symetries Qantiques,
Proc.Les-Houches summer school, LXIV.
Eds. A.Connes, K.Kawedzki, J.Zinn-Justin. North-Holland, 1998, 149--211,
hep-th/9605187.

\bibitem{F95}
  L.~D.~Faddeev,
 {\em Discrete Heisenberg-Weyl group and modular group},
  Lett. Math. Phys.  {\bf 34} (1995), 249--254; hep-th/9504111.

\bibitem{F99} L.~D.~Faddeev,
{\em Modular double of a  quantum group}, Conf. Mosh\'e Flato 1999,
vol. I, Math. Phys. Stud. {\bf 21}, Kluwer, Dordrecht, 2000, pp. 149--156;
math.QA/9912078.

\bibitem{FKV}
  L.~D.~Faddeev, R.~M.~Kashaev, A.~Y.~Volkov,
{\em Strongly coupled quantum discrete Liouville theory. 1. Algebraic approach and duality},
  Commun. Math. Phys. {\bf 219} (2001), 199--219; hep-th/0006156.
				
\bibitem{FTT83}
 V.~O.~Tarasov, L.~A.~Takhtajan and L.~D.~Faddeev,
{\em Local Hamiltonians for integrable quantum models on a lattice},
Teor. Mat. Fiz. {\bf 57} (1983), 163--181
(Theor. Math. Phys.  {\bf 57} (1983), 1059--1073).

\bibitem{VF}
A.~Yu.~Volkov and L.~D.~Faddeev, {\em Yang-Baxterization of the
quantum dilogarithm}, Zapiski POMI {\bf 224} (1995), 146--154
(J. Math. Sciences {\bf 88} (2) (1998), 202--207).

\bibitem{Had}
  L.~Hadasz, M.~Pawelkiewicz, V.~Schomerus,
  {\it Self-dual Continuous Series of Representations for $\mathcal{U}_q(sl(2))$ and $\mathcal{U}_q(osp(1|2))$},
  JHEP {\bf 1410} (2014) 91
  [arXiv:1305.4596 [hep-th]].

\bibitem{Jim}
M. Jimbo (ed), {\it Yang-Baxter equation in integrable systems},
Adv. Ser. Math. Phys., 10, World Scientific (Singapore), 1990.		

\bibitem{KhT}
S.~M.~Khoroshkin, V.~N.~Tolstoy,
{\it Yangian Double}, Lett. Math. Phys.
{\bf 36} (1996), 373--402; hep-th/9406194.

\bibitem{Khoroshkin:2014hla}
  S.~Khoroshkin and Z.~Tsuboi,
  {\it The universal R-matrix and factorization
  of the L-operators related to the Baxter Q-operators},
  J.\ Phys.\ A {\bf 47}, 192003 (2014)
  [arXiv:1401.0474 [math-ph]].

\bibitem{KrZa97}
I.~Krichever and A.~Zabrodin, {\em Vacuum curves of elliptic
$L$-operators and representations of Sklyanin algebra},
Amer. Math. Soc. Transl. Ser. {\bf 2}, Vol. {\bf 191} (1999), 199--221;
solv-int/9801022.

\bibitem{KuSk}
P.~P.~Kulish, E.~K.~Sklyanin, {\em Solutions of the Yang-Baxter equation},
Zapiski LOMI {\bf 95} (1980), 129--160
(J. Soviet Math. {\bf 19} (5) (1982), 1596--1620).


\bibitem{KRS81}
  P.~P.~Kulish, N.~Y.~Reshetikhin and E.~K.~Sklyanin,
  {\em Yang-Baxter Equation and Representation Theory. 1.},
  Lett. Math. Phys.  {\bf 5} (1981), 393--403

\bibitem{KuSk1}
P.~P.~Kulish and E.~K.~Sklyanin,~{\it Quantum spectral transform
method. Recent developments}, Lect. Notes in
Physics, {\bf 151} (1982), 61--119.

\bibitem{Mangazeev:2014gwa}
  V.~V.~Mangazeev,
  {\em On the Yang-Baxter equation for the six-vertex model},
  Nucl.\ Phys.\ B {\bf 882} (2014) 70
  [arXiv:1401.6494 [math-ph]].

\bibitem{Mangazeev:2014bqa}
  V.~V.~Mangazeev,
  {\em $Q$-operators in the six-vertex model},
  Nucl.\ Phys.\ B {\bf 886}, 166 (2014)
  [arXiv:1406.0662 [math-ph]].

\bibitem{Pawelkiewicz:2013wga}
  M.~Pawelkiewicz, V.~Schomerus and P.~Suchanek,
{\it The universal Racah-Wigner symbol for U$_q$(osp(1|2))},
  JHEP {\bf 1404} (2014) 079
  [arXiv:1307.6866 [hep-th]].
		
\bibitem
{Rains} E.~M.~Rains,
{\it $BC_n$-symmetric abelian functions}, Duke Math. J.
{\bf  135} (1) (2006), 99--180.

\bibitem
{ros:elementary} H.~Rosengren,
{\em An elementary approach to $6j$-symbols (classical, quantum, rational,
trigonometric, and elliptic)}, Ramanujan J. {\bf 13} (2007), 131--166; math/0312310.

\bibitem
{ros:sklyanin} H.~Rosengren,
{\em Sklyanin invariant integration},
Internat. Math. Res. Notices, no. {\bf 60} (2004), 3207--3232; math/0405072.
	
\bibitem{rui}
S.~N.~M.~Ruijsenaars, {\em First order analytic difference equations and integrable quantum systems},
J. Math. Phys. {\bf 38} (1997), 1069--1146.

\bibitem
{skl1} E. K. Sklyanin, {\em  Some algebraic structures connected with
the Yang-Baxter equation}, Funkz. Analiz i ego Pril. {\bf 16} (4) (1982),  27--34.

\bibitem{skl2} E. K. Sklyanin, {\it On some algebraic structures
related to Yang-Baxter equation: representations of the quantum algebra},
Funkz. Analiz i ego Pril. {\bf 17} (1983), 34--48.

\bibitem{AA2008}
V.~P.~Spiridonov,
{\em Continuous biorthogonality of the elliptic hypergeometric function},
Algebra i Analiz {\bf 20} (5) (2008), 155--185
(St. Petersburg Math. J. {\bf 20} (5) (2009) 791--812), arXiv:0801.4137 [math.CA].

\bibitem{spi:umn}
V. P. Spiridonov,
\textit{On the elliptic beta function}, Uspekhi Mat. Nauk {\bf 56} (1)
(2001), 181--182 (Russian Math. Surveys {\bf 56} (1) (2001),
185--186).

\bibitem{spi:bailey}
V. P. Spiridonov, \textit{A Bailey tree for
integrals}, Teor. Mat. Fiz. {\bf 139} (2004), 104--111 (Theor. Math.
Phys. {\bf 139} (2004), 536--541), math.CA/0312502.


\bibitem{spi:essays}
V. P. Spiridonov,
{\em Essays on the theory of elliptic
hypergeometric functions}, Uspekhi Mat. Nauk {\bf 63} (3)
(2008), 3--72 (Russian Math. Surveys {\bf 63} (3) (2008), 405--472),
arXiv:0805.3135 [math.CA].

\bibitem{spi-war:inversions}
V. P. Spiridonov and S. O. Warnaar, \textit{
Inversions of integral operators and elliptic beta integrals on root
systems}, Adv. Math. {\bf 207} (2006), 91--132.
	
\bibitem{Vol05}
  A.~Y.~Volkov,
{\em Noncommutative hypergeometry},
  Commun. Math. Phys.  {\bf 258} (2005), 257--273; math.QA/0312084.


\bibitem{Z}
A.~Zabrodin, {\em On the spectral curve of the difference Lame operator},
Int. Math. Research Notices, no. 11 (1999), 589--614; arXiv:math/9812161.

\end{thebibliography}
\end{document}